\documentclass[a4paper,12pt]{elsarticle}
\usepackage{graphicx}
\usepackage{hologo}
\usepackage{amsmath}
\usepackage{caption}
\usepackage{verbatim} 
\usepackage[table,xcdraw]{xcolor}
\usepackage[utf8]{inputenc}
\usepackage{hyperref}
\usepackage{float}
\usepackage[section]{placeins}
\usepackage{lineno}
\usepackage{orcidlink}

\begin{document}
\begin{frontmatter}

\title{Multidisciplinary Design Optimization for Wave-Driven Desalination Systems}

\author{\corref{cor1}Nate DeGoede\orcidlink{0000-0002-6226-0647}$^{a}$}
\author{Maha N. Haji\orcidlink{0000-0002-2953-7253}$^a$}

\affiliation[a]{Department of Mechanical Engineering, University of Michigan, Ann Arbor, MI 48109, USA}

\begin{abstract}
Wave-driven desalination systems are an innovative solution to the global freshwater crisis, leveraging the complementary characteristics of seawater reverse osmosis and wave energy converters. However, the high costs of this system pose a significant barrier to widespread adoption. Optimization can help these systems reach a more competitive levelized cost of water, but the highly coupled nature of the system necessitates a multidisciplinary design optimization approach. This paper presents a holistic, multidisciplinary design optimization framework for wave-driven desalination system design, integrating models for wave energy converter hydrodynamics, power take-off transmission, seawater reverse osmosis constraints, and economic analysis. This study demonstrates the impact of multidisciplinary design optimization for wave-driven desalination systems, resulting in a 69.5\% reduction in levelized cost of water within this modeling framework compared to a nominal design. We demonstrate that multidisciplinary design optimization outperforms two different sequential design approaches, yielding lower levelized costs of water and substantially different optimal designs. The multidisciplinary design optimization results suggest major design changes compared to designs found in the literature. Notably, smaller wave energy converters and larger pistons, along with smaller accumulators and larger seawater reverse osmosis plant installations, are preferred within this modeling framework. These design trends are consistent across a range of sea states, suggesting potential generalizability beyond a single location. This study demonstrates the importance of holistic modeling and co-design for wave-driven desalination systems and establishes an effective optimization framework for future studies to build upon.

\end{abstract}

\cortext[cor1]{Corresponding author \\ \textit{Email address: degoeden@umich.edu}}

\end{frontmatter}
\section{Introduction}
The global freshwater crisis necessitates innovative solutions for clean water production. Demand for freshwater is anticipated to grow by over 40\% by 2050, while droughts, urbanization, and uneven water distribution further strain existing supplies~\cite{watershortage2015}. Seawater reverse osmosis (SWRO) is a proven method for desalination, but is energy intensive, typically requiring 2-4 kWh/m$^3$ of freshwater produced~\cite{Li2018}. This results in significant energy costs, accounting for 25\%-40\% of typical total freshwater production expenses~\cite{blue_econ}. When powered by fossil fuels, it further exacerbates global water scarcity challenges~\cite{nytdrought}. Ensuring sustainability, therefore requires SWRO powered efficiently through renewable energy sources.

Various renewable energy sources, such as wind~\cite{Lai2016} and solar~\cite{Elkadeem2024}, have been explored for powering SWRO. Wave energy converters (WECs) are a particularly promising option due their natural co-location with coastal SWRO plants. Beyond sharing a marine environment, WECs and SWRO plants have complementary operational characteristics~\cite{blue_econ}. SWRO requires high-pressure seawater, which WECs can directly supply via hydraulic power take-off (PTO) systems, enabling wave-driven desalination~\cite{Davies2005}. This direct-drive approach avoids intermediate electricity generation, potentially improving overall system efficiency and reducing system complexity. Despite wind and solar energy's greater level of technical maturity and lower costs, this potential direct-drive approach makes wave powered SWRO worth studying. 

Despite their potential, WECs face significant challenges limiting their widespread adoption~\cite{Falcao2010}. These challenges include maintaining PTO efficiency over a broad operating spectrum, ensuring survivability and long-term reliability in harsh marine environments, and reducing high system costs, among many others. While survivability and reliability remain active areas of materials and structural research, design optimization can be leveraged to improve system efficiency and economic viability. Multidisciplinary design optimization (MDO) offers a systematic framework for addressing this challenge by considering multiple subsystems simultaneously and optimizing overall performance rather than isolated components~\cite{Sobieski}. While widely used in industries such as aerospace, MDO has only recently gained traction in the WEC community. 

Control co-design (CCD), a subset of MDO that jointly optimizes system design and control strategy, has received growing attention in this field. Peña-Sanchez et al. (2022) used CCD to show that PTO sizing is a non-linear design problem~\cite{PenaSanchez2022}. Rosati and Ringwood (2023) applied CCD to an oscillating water column WEC and found PTO sizing to be crucial for minimizing levelized cost of energy (LCOE)~\cite{Rosati2023}. Michelén Ströfer et al. (2023) achieved a 22\% increase in electrical power output using CCD, demonstrating the importance of impedance matching for WEC PTOs~\cite{Stroefer2023}. Grasberger et al. (2024) reported a 60\% reduction in LCOE using MDO when compared to traditional design methods~\cite{Grasberger2024}. 

MDO is particularly effective for these problems due to strong interactions between subsystems. Garcia-Rosa and Ringwood (2016) show that LCOE-optimal WEC and PTO design highly sensitive to control strategy, underscoring the importance of CCD~\cite{GarciaRosa2016}. Coe et al. further explored these interactions through impedance matching studies. Coe et al. (2021) demonstrated that while well-established control techniques, (e.g., causal feedback control and impedance matching), are effective for WEC control, they depend on a well-designed PTO system that properly shapes system impedance~\cite{Coe2021}. Coe et al. (2025) extended this work by applying impedance matching directly to PTO design~\cite{Coe2025}. Although these studies focus on rack-and-pinion style WEC PTOs for electricity generation, impedance matching principles are also relevant to hydraulic PTOs with rectifying behaviors~\cite{Ladan2015}, such as those used in wave-driven desalination systems.

These MDO and impedance matching studies highlight the need for holistic models that incorporate wave dynamics, PTO performance, and control strategies in optimizing electricity-generating WECs. However, these approaches have not yet been extended to wave-driven desalination systems, leaving a critical gap in research.

Ylänen and Lampinen (2014) explored pump sizing and operating pressures in wave-driven desalination systems, but did not account for transient dynamics or the resulting impact on WEC performance~\cite{Ylaenen2014}. Yu and Jenne (2017) conducted a techno-economic analysis of desalination plant capacity sizing trade-offs, but considered this decision in isolation, limiting the completeness of the trade-off exploration~\cite{YJecon2017}. Their follow-up work (2018) modeled additional PTO components such as accumulators and pressure relief valves, but only in binary terms (either included or not), without optimizing their sizing~\cite{Yu2018}. Brodersen et al. (2022) compared PTO architectures for a batch process wave-driven desalination system, but did not perform design optimization or sizing studies~\cite{brodersen2022}. Suchithra et al. (2022) carried out sequential sizing, but without capturing system interactions in the design process \cite{Suchithra2022}. Simmons and Van de Ven (2023) examined how PTO architecture affects freshwater production and flow smoothing, but focused solely on PTO design without co-optimizing the WEC design~\cite{Simmons2023}. Although these papers present holistic models that account for the coupled dynamics of wave-driven desalination systems, they do not leverage them within an MDO framework.

When optimizing WEC systems, it is important to account for sea state variability. Robertson et al. (2022) showed that optimal WEC geometry depends strongly on the incident wave spectrum~\cite{Robertson2022}. Because wave conditions vary even at a single location, designs should be evaluated across a range of sea states. Many studies adopt this approach; for example, Stroefer et al. (2023) use a weighted average of performance across multiple sea states as the objective function~\cite{Stroefer2023}. In contrast, this study optimizes the design for various sea states individually, providing clearer insight into how sensitive of the optimal design is to sea state. 


This paper addresses these gaps by introducing a holistic multidisciplinary design optimization framework for wave-driven desalination system design. While MDO is well established and has been applied broadly in energy systems, prior applications to wave energy have largely focused on electricity-generating WECs, and prior wave-driven desalination studies have not yet leveraged MDO. This paper formulates the wave-driven desalination system as a coupled co-design problem in which WEC hydrodynamics, power take-off transmission, seawater reverse osmosis operating constraints, and economic performance are optimized simultaneously. This coupling is important because design choices that improve wave energy capture do not necessarily minimize desalinated water cost once PTO behavior, RO feasibility constraints, and sea-state-dependent operation are included. The core contributions of this work are therefore: (i) an integrated MDO formulation and modeling workflow for system-level co-design of wave-driven desalination systems; (ii) a quantitative demonstration of the value of simultaneous co-optimization relative to more sequential or decoupled design approaches; and (iii) identification of robust design trends across a range of representative sea states, providing insight into how optimal WDDS configurations depend on wave-resource conditions. In Section~\ref{sec:problem}, the formal MDO problem formulation is presented, defining the design space and providing an overview of the holistic model. Section~\ref{sec:model} details the disciplinary models. Section~\ref{sec:results} presents the optimization study results, which are further discussed in Section~\ref{sec:discussion}. Finally, conclusions and study limitations are presented in Section~\ref{sec:conclusion}.

\section{Problem Formulation} \label{sec:problem}
\begin{figure}[b!]
    \centering
    \includegraphics[width=0.6\linewidth]{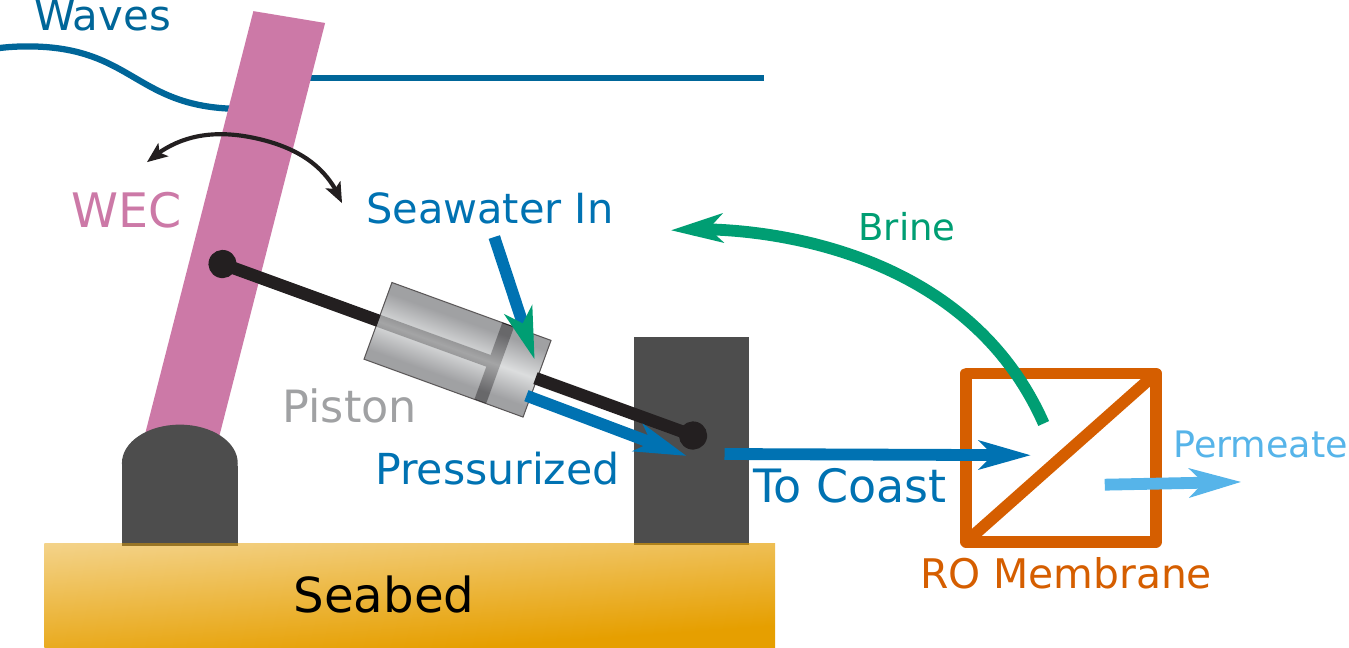}
    \captionof{figure}{Simple wave-driven desalination system concept sketch.}
    \label{fig:WDDS}
\end{figure}

Figure~\ref{fig:WDDS} illustrates a simplified wave-driven desalination system, where an oscillating surge wave energy converter (OSWEC) drives a hydraulic cylinder to pressurize seawater and supply an onshore SWRO desalination plant, producing freshwater without significant external energy input. 
\begin{figure}[b!]
    \centering
    \includegraphics[width=0.5\linewidth]{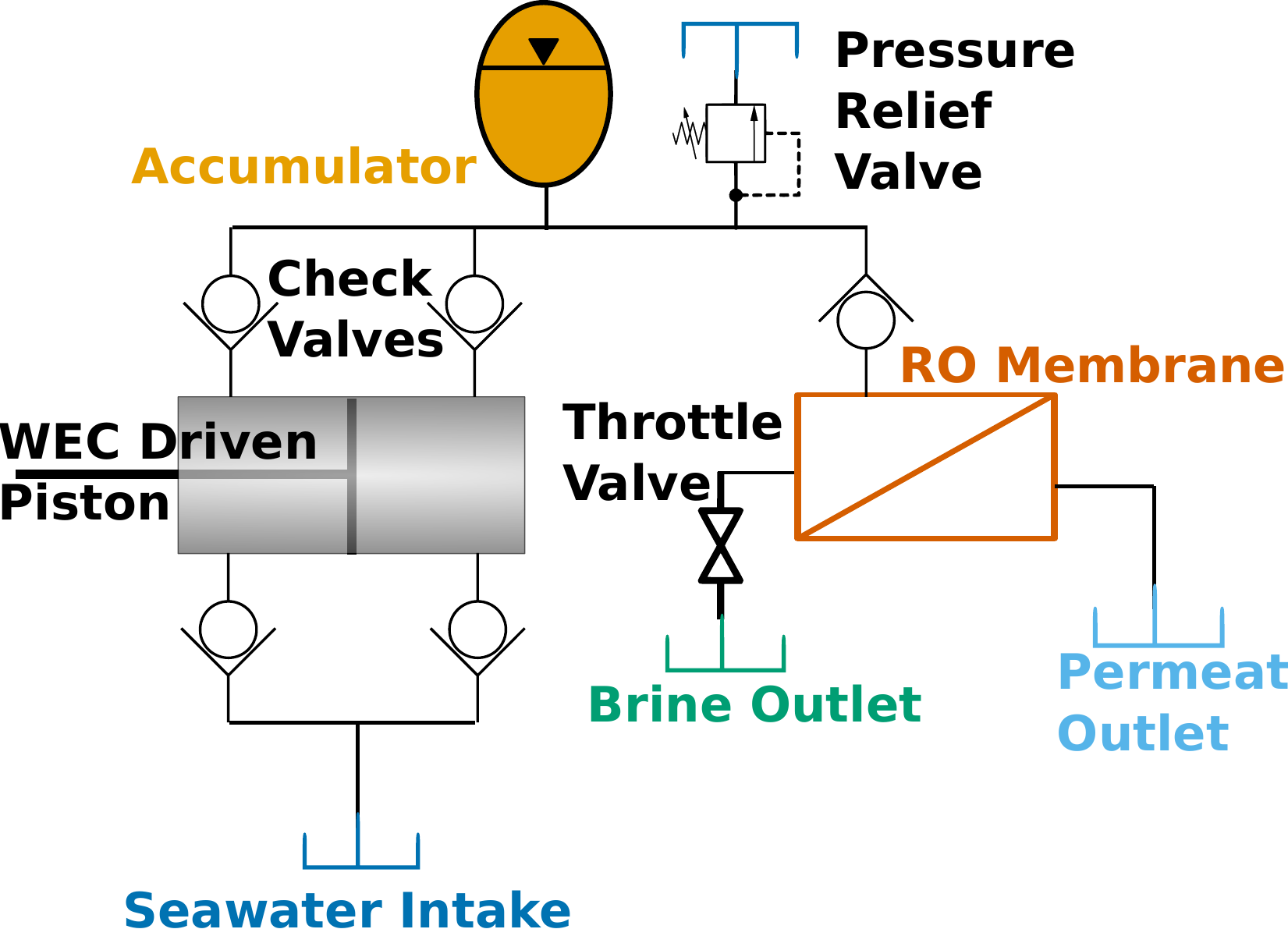}
    \captionof{figure}{Hydraulic circuit diagram of the wave-driven desalination system.}
    \label{fig:hydraulics}
\end{figure}
While Figure~\ref{fig:WDDS} provides a high-level overview, it omits several hydraulic components included in this study. Figure~\ref{fig:hydraulics} presents the full hydraulic circuit, which incorporates an accumulator for pressure smoothing over short time-scales, a pressure relief valve to prevent exceeding SWRO plant capacity, a brine throttle valve to regulate recovery ratio, and directional valves. Mi et al. (2023) experimentally tested a similar system, demonstrating feasibility, although their system did not include a pressure relief valve~\cite{Mi2023}. While alternative configurations are possible, such as the use additional storage in place of a pressure relief valve or the incorporation of energy recovery devices to recycle energy from the brine stream, the present configuration was selected as a relatively simple platform for evaluating representative design trade-offs in wave-driven SWRO systems. In addition, this work does not consider site selection and instead assumes a deployment site that satisfies the bathymetry, water depth, and distance-from-shore constraints associated with a nearshore fixed-bottom installation.

\subsection{Multidisciplinary Design Optimization Framework}
The optimization problem in this study is formally defined as:
\begin{align*} 
    \text{Minimize} \quad & \text{LCOW}(\mathbf{x},\mathbf{p}; \hat{u})  \\
    \text{by varying} \quad & \mathbf{x}\\
    \text{subject to} \quad &  \mathbf{g} \leq 0 \\
    \text{while solving} \quad & \mathbf{R}(\mathbf{x},\mathbf{p};\hat{u}) = 0\\
    \text{for} \quad & \hat{u}
    \label{eq:problem}
\end{align*}
\noindent where LCOW (levelized cost of water) is the objective function (discussed in Section~\ref{sec:econ}). The design vector, $\mathbf{x}$, contains variables related to WEC, PTO, and SWRO plant, summarized in Table~\ref{tab:design_space}. $\mathbf{p}$ contains parameters related to environmental conditions and fixed design choices (shown in \ref{app:params}). $\hat{u}$ represents the coupled variables solved for by $\mathbf{R}$, the governing equations of the various disciplinary modules: geometry, seawater desalination, hydrodynamics, rigid body dynamics, hydraulics, and economics. The problem is subject to inequality constraints ($\mathbf{g}$) on piston motion and pressure, enforced within the system dynamics module (rigid-body dynamics and hydraulics submodules), which is further described in Section~\ref{sec:sysdyn}.

\begin{figure*}[t!]
    \centering
    \includegraphics[width=\linewidth]{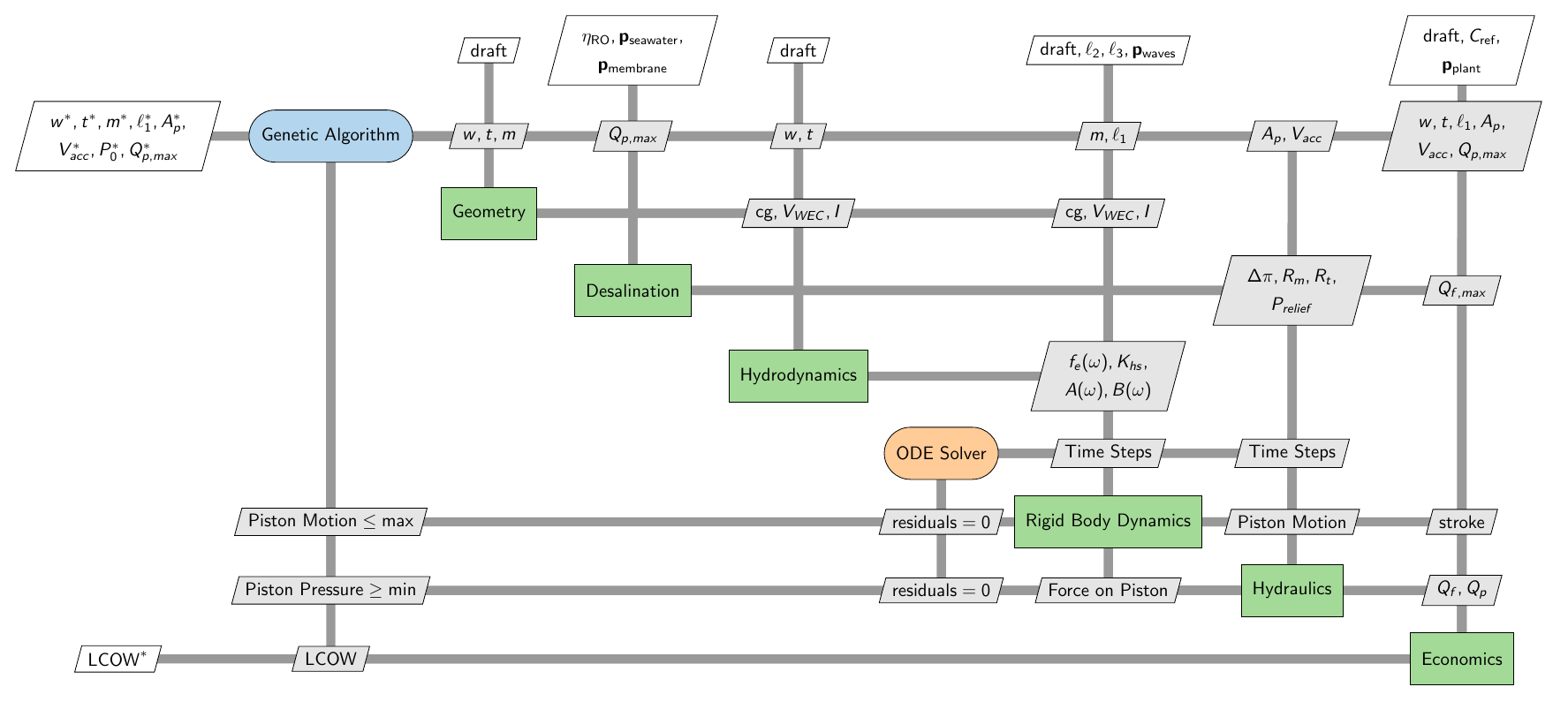}
    \captionof{figure}{xDSM diagram of the wave-driven desalination system optimization problem. Note $\mathbf{p}_\text{seawater}$, $\mathbf{p}_\text{membrane}$, $\mathbf{p}_\text{waves}$, and $\mathbf{p}_\text{plant}$ represent vectors of parameters related to the membrane, seawater, sea state, and desalination plant architecture respectively.}
    \label{fig:xdsm}
\end{figure*}

Figure~\ref{fig:xdsm} presents a graphical representation of the MDO problem using an Extended Design Structure Matrix (xDSM)~\cite{xdsm}. The diagram uses gray lines to illustrate the flow of information between the disciplinary modules (green boxes), optimizer (blue stadium), and internal solver (orange stadium), while gray parallelograms indicate module inputs and outputs. The white parallelograms show optimizer level inputs and outputs.

\begin{figure}[b!]
    \centering
    \includegraphics[width=0.8\linewidth]{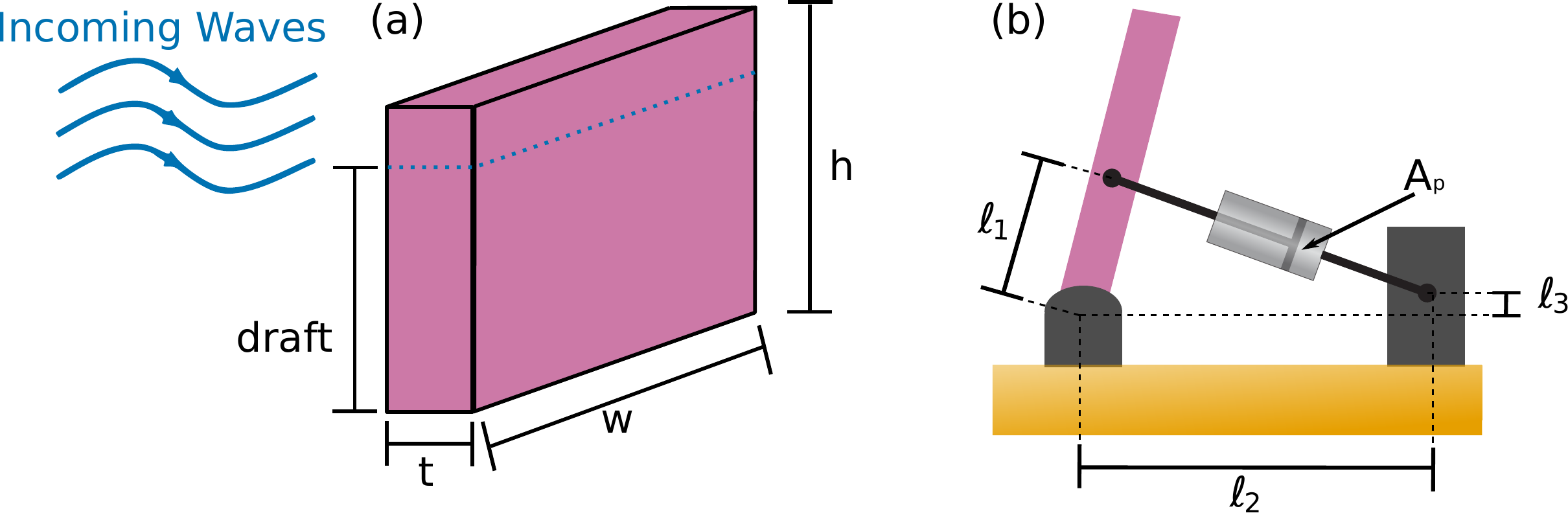}
    \captionof{figure}{WEC (a) and mechanism (b) dimensions. Note that $h$ represents the WEC height, which is a parameter rather than a design variable.}
    \label{fig:dvs}
\end{figure}

A summary of the design space is shown in Table~\ref{tab:design_space}. WEC variables consist of the width, $w$ [m], thickness, $t$ [m], and mass, $m$ [kg], of the OSWEC. PTO variables include the OSWEC hinge to PTO joint length, $\ell_1$ [m], piston area, $A_p$ [m$^2$], accumulator volume, $V_{acc}$ [m$^3$], and accumulator pre-charge pressure, $P_0$ [MPa]. Finally, the SWRO plant variable is the plant capacity, $Q_{p,max}$ [m$^3$/day]. Many of these variables can be visualized in Figure~\ref{fig:dvs}.
\begin{table}[t!]
    \centering
    \caption{Design Variables, including nominal values \cite{Yu2018}. Bounds with citations use the same bounds as the cited work, otherwise, bounds are set by engineering judgment.}
    \begin{tabular}{cccc}
        \hline
        Variable & Nominal Value & Lower Bound & Upper Bound\\
        \hline
        $w$ & 18 m & 4 m \cite{Grasberger2024} & 24 m \\
        $t$ & 2 m & 0.8 m \cite{Grasberger2024} & 3 m \\
        $m$ & 127$\times10^3$ kg & 50$\times10^3$ kg & 500$\times10^3$ kg \\
        $\ell_1$ & 2 m & 0.1 m & 4 m \\
        $A_p$ & 0.26 m$^2$ & 0.1 m$^2$ & 1 m$^2$ \\
        $V_{acc}$ & 4 m$^3$ & 0.01 m$^3$ & 6 m$^3$ \\
        $P_0$ & 3 MPa & 3 MPa & 6 MPa \\
        $Q_{p,max}$ & 3150 m$^3$/day & 1000 m$^3$/day & 10000 m$^3$/day\\
        \hline
    \end{tabular}
    \label{tab:design_space} 
\end{table}

Some design parameters are excluded from the design space. WEC height, draft and mechanism variables $\ell_2$ and $\ell_3$ are fixed to reduce the dimensionality of the design space dimensionality, since the remaining variables provide sufficient control of the system's dynamic characteristics. WEC thickness adjusts the stiffness of the WEC with minimal impact on other dynamic properties, while WEC mass controls the WEC's inertia. Although width affects WEC stiffness, inertia, and excitation force, stiffness and inertia can already be tuned independently through thickness and mass, aldescriptionlowing width to primarily adjust the excitation force and damping. This combination provides sufficient control over the WEC's dynamic response. The mechanism variables only affect the effective mechanical advantage of the pumping mechanism; therefore, only the most sensitive parameter, $\ell_1$, is included in the design space.

This study uses parameters similar to those used by Yu and Jenne (2018) \cite{Yu2018}. Full tables of parameters, including environmental conditions, fixed design variables, and economic parameters, can be found in \ref{app:params}. Constraints are applied to device stroke length, maximum hydraulic circuit pressure, and minimum piston cylinder pressure. These are all enforced in the system dynamics module, Section ~\ref{sec:sysdyn}.

Due to the absence of accessible gradients in the hydrodynamics and system dynamics modules, a genetic algorithm (GA) is utilized. Table~\ref{tab:paramsga} contains the GA hyperparameters used. Since premature convergence was a major issue in the optimization process, an aggressive immigration strategy is used, where every 50 generations, 75\% of the population is replaced with new randomly generated individuals. This strategy is effective at preventing early convergence, but it increases the computational cost of the optimization process. 
\begin{table}[b]
    \centering
    \caption{Genetic algorithm hyperparameters}
    \begin{tabular}{ccc}
        \hline
        \textbf{Parameter} & \textbf{Value} & \textbf{Units} \\
        \hline
        population size & 400 & - \\
        mutation rate & 0.2 & - \\
        crossover rate & 0.8 & - \\
        number of elites kept & 1 & - \\
        tournament size & 2 & - \\
        bits per variable & 8 & - \\
        immigration interval & 50 & generations \\
        immigrant population size & 300 & - \\
        \hline
    \end{tabular}
    \label{tab:paramsga}
\end{table}
\subsection{Sequential Design Optimization Approach} \label{sec:sdo}
\begin{figure}[t!]
    \centering
    \includegraphics[width=0.85\linewidth]{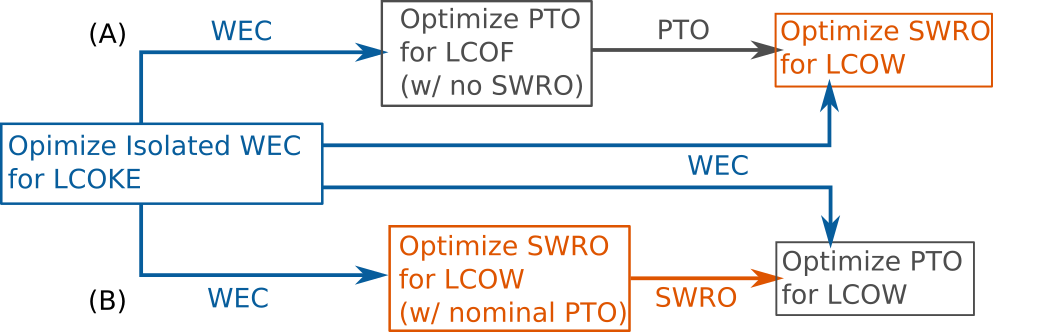}
    \captionof{figure}{Flow charts for the sequential design optimization approaches.}
    \label{fig:sdo_flow}
\end{figure}

MDO can be a computationally intensive task, which limits its potential applications. To quantify the importance of MDO for this specific application, a comparison is made between MDO and two sequential design optimization (SDO) approaches, shown in Figure~\ref{fig:sdo_flow}. In SDO, the design of each subsystem (WEC, PTO, and SWRO plant) is optimized sequentially. The WEC is optimized first, as this reflects the trends of design studies in WEC literature, notably Suchithra et al. (2022), who do this for a wave-driven desalination system \cite{Suchithra2022}. To optimize the WEC design, we optimize for the levelized cost of WEC kinetic energy (LCOKE) [\$/kWh] when excited at the peak period and disconnected from PTO. The only costs associated with this are the WEC costs.
To optimize the PTO and SWRO plant designs, we take two approaches. The first approach (A) optimizes the PTO for levelized cost of flow (LCOF) [\$/m$^3$] with the SWRO plant (membrane and pressure relief valve) replaced by a throttle valve. This results in optimizing the PTO to deliver the maximum feed flow into the SWRO plant at the lowest PTO cost. This ``optimized'' PTO is then used in the optimization of the SWRO plant, which optimizes the SWRO plant design for LCOW.

The second approach (B) chooses a nominal PTO design from literature~\cite{Yu2018} and optimizes the SWRO plant design for LCOW using this fixed PTO design. This optimal SWRO plant design is then fixed while optimizing the PTO design for LCOW.

\subsection{Sensitivity Analysis} \label{sec:sensitivity}
The optimal design of the wave-driven desalination system is expected to change with different sea states. Since one of the goals of this study is to explore areas of potential design improvement through MDO, it is therefore important to see if the same design trends are recommended in a variety of sea states. To this end, optimizations are performed for 20 different sea states, which can be found in \ref{app:seastates}.

To balance representativeness and computational tractability, the representative sea states were defined as the centroids of 20 sea-state clusters obtained using K-means clustering~\cite{Arthur2007}. Hourly significant wave heights and peak periods observations from 2015-2024 are collected from the National Oceanographic and Atmospheric Administration (NOAA) National Data Buoy Center (NDBC) \cite{ndbc} for five buoys spanning a different regions: Guam (52200), Hilo HI (51206), Santa Monica Bay (46221), San Juan PR (41053), and Georges Bank (44011). For each location, K-means clustering is performed on the significant wave height and peak period data to create 10 representative sea states for the region. These 50 representative sea states are then clustered again using K-means clustering to create the 20 final sea states used in the sensitivity study. For simplicity, we do not consider differences in bathymetry, depth, distance to shore, or different spectra shapes that these different locations might include. While these would be important considerations in a siting study, this sensitivity study is more just interested in understanding the impact different sea states could have on the MDO optimal design.

\section{System Modeling} \label{sec:model}
Building upon DeGeode and Haji (2025)~\cite{idetc2025}, the xDSM in Figure~\ref{fig:xdsm} shows the five primary disciplinary modules: Geometry, which maps WEC design variables to geometric properties; Desalination, which calculates the system dynamic properties of the SWRO plant; Hydrodynamics, which determines various hydrodynamic coefficients; System Dynamics, which solves the coupled dynamics of the wave-driven desalination system; and Economics, which calculates the LCOW of the system. The Geometry module is straightforward and not discussed further. The remaining four modules are detailed in the following subsections.

\subsection{Desalination Module} \label{sec:desal}
The desalination module determines key SWRO plant parameters for a single-stage plant. It considers the capacity of the plant in addition to membrane parameters and seawater composition. The governing equation for the reverse osmosis (RO) process is:
\begin{equation}
    \label{eq:ro}
    Q_p = A_w A_m (\Delta P - \Delta \pi),
\end{equation}
\noindent where $Q_p$ [m$^3$/s] is the permeate flow rate, $A_w$ [m$^3$/N-s] is the water permeability coefficient, $A_m$ [m$^2$] is the total membrane area, $\Delta P$ [Pa] is the pressure difference across the membrane, and $\Delta \pi$ [Pa] is the osmotic pressure difference across the membrane. This osmotic pressure difference is calculated using:
\begin{equation}
    \label{eq:osmotic_pressure}
    \pi = iCRT,
\end{equation}
\noindent where $i$ [-] is the number of ions produced per molecule of solute, $C$ [mol/m$^3$] is the concentration of the solute, $R$ [J/K-mol] is the ideal gas constant, and $T$ [K] is the temperature \cite{separationprocesses}. For simplicity, $\Delta \pi$ is set as the osmotic pressure of the seawater minus the osmotic pressure of the target permeate concentration. A better model would consider the time variant concentrations on both sides of the membrane, but is outside the scope of this study.

The water permeability coefficient ($A_w$) is set to a constant value, matching the work by Yu and Jenne (2018)~\cite{Yu2018}. Sitterley et al. (2025) found that using a water permeability coefficient that varies with feed condition improves the accuracy of transient SWRO models~\cite{Kurby2025}, but at the fidelity level of this study, a constant value is sufficient. This constant value is dependent upon membrane selection; for this paper, we use the same membrane as Yu and Jenne (2018), the SW30HR-380 Dry from DuPont~\cite{SW30HR380}. Membrane area $A_m$ is set by calculating the amount of SW30HR-380 area required to meet the SWRO plant capacity design variable, $Q_{p,max}$:
\begin{equation}
    A_m = \frac{Q_{p,max}}{Q_0}A_0,
    \label{eq:membrane_area}
\end{equation} 
\noindent where $Q_0$ and $A_0$ are the single-unit production and membrane area for the SW30HR-380 membrane, found on the datasheet~\cite{SW30HR380}, and included in Table~\ref{tab:paramsswro}.

For convenience a membrane resistance ($R_m$ [MPa-s/m$^3$]) is defined as: 
\begin{equation}
    R_m = \frac{1}{A_w A_m}
    \label{eq:membrane_resistance}
\end{equation}
\noindent this is then used to determine the pressure relief valve set point, $P_{\text{relief}}$ [MPa]:
\begin{equation}
    P_{\text{relief}} = Q_{p,max}R_m + \Delta \pi
    \label{eq:pressure_relief}
\end{equation}
\noindent This equation ensures that if the flow exceeds the plant capacity, the pressure relief valve will open.

To force flow to pass through the membrane, resistance is required on the brine side. For this study, a simple throttle valve provides this resistance, but an ERU could also be used. The resistance of the throttle valve $R_t$ [MPa-s/m$^3$] is calculated as: 
\begin{equation}
    R_t = \frac{P_{\text{relief}}}{Q_{p,max}(\frac{1}{\eta_{RO}} - 1)}
\end{equation}
\noindent where $\eta_{RO}$ [-] is the recommended recovery ratio of the SW30HR-380 membrane given the seawater composition, as determined using WAVE, Dow Chemical Company's water-treatment process design tool~\cite{wave}. This equation ensures that when operating at full capacity, the recommended recovery ratio is used. When operating below capacity, the recovery ratio will be lower. Although a controlled brine valve could be implemented to regulate either recovery ratio or pressure more directly~\cite{Bacelli2009}, a fixed valve configuration is adopted in this study to maintain model simplicity.

Beyond the assumptions of constant membrane permeability and osmotic pressure difference, the desalination model neglects concentration polarization, fouling, and membrane degradation under unsteady operating conditions. While these phenomena may affect system performance and longevity, their exclusion is consistent with the level of fidelity typically employed in systems-level analyses. Among these effects, membrane degradation under unsteady loading may be particularly important, although additional research is needed to establish its magnitude and governing mechanisms~\cite{Sitterley2022}.

\subsection{Hydrodynamics} \label{sec:hydro}
Falnes and Kurniawan (2020) define the oscillating WEC equation of motion as~\cite{falnes2020}:
\begin{equation}
    \label{eq:wec}
    I\ddot{\xi} = f_e - f_r - f_b - f_v - f_f - f_u  
\end{equation}
\noindent where $I$ [m or kg-m$^2$] is the inertia of the WEC, $\xi$ [m or rad] is the body motion, $f_e$ [N or Nm] is the excitation force/moment due to incident waves, $f_r$ [N or Nm] is the radiation force/moment due to the WEC’s oscillations, $f_b$ [N or Nm] is the hydrostatic force/moment balancing buoyancy and weight, $f_v$ [N or Nm] is the drag force/moment arising from nonlinear viscous effects, $f_f$ [N or Nm] captures the friction forces/moments, and $f_u$ [N or Nm] is the PTO force/moment, representing the force applied by the WEC's power conversion system. The PTO force/moment ($f_u$) is modeled in the system dynamics module, Section~\ref{sec:sysdyn}. This study assumes linear potential flow (linear, irrotational, inviscid flow) and omits the drag ($f_v$) and friction effects ($f_f$) from the WEC equation of motion. This simplification is also adopted by Grasberger et al. (2024)~\cite{Grasberger2024} and is prevalent in the wave energy converter optimization literature, despite known limitations when modeling large-amplitude motions.

The hydrostatic force/moment accounts for the combination of buoyancy and weight and is defined as:
\begin{equation}
    \label{eq:hydrostatic}
    f_{b} = K_{hs} \xi,
\end{equation}
\noindent where $K_{hs}$ [N/m or Nm/rad] is the hydrostatic stiffness matrix. The radiation force/moment accounts for the waves generated by the WEC's motion and is defined by the Cummins equation \cite{wecsim}, which accounts for fluid memory effects:
\begin{equation}
    \label{eq:radiation}
    f_r(t) = -A_\infty \ddot{\xi}(t) - \int_0^t \left[ \frac{2}{\pi}\int_0^\infty B(\omega)cos(\omega t -\omega\tau)d\omega \right] \dot{\xi}(\tau)d\tau,
\end{equation}
where $A_\infty$ [kg or kg-m$^2$] is the added mass matrix at infinite frequency and $B(\omega)$ [N-s/m or Nm-s/rad] is the radiation damping matrix, a function of frequency, $\omega$ [rad/s]. The excitation force/moment accounts for the force/moment on the WEC due to the incident waves and is defined as:
\begin{equation}
    f_e = F_e(\omega)\exp\left({i\omega t}\right) A_{wave},
\end{equation}
\noindent where $F_e(\omega)$ [N/m or Nm/m] is the excitation force/moment amplitude operator and $A_{wave}$ [m] is the wave amplitude.

In this study, we will use irregular waves, modeled in the rigid body dynamics module discussed in Section~\ref{sec:sysdyn}. The wave amplitudes are determined from the sea state, in this study, we use Pierson-Moskowitz wave spectra, which are described by the following equation:
\begin{equation}
S(\omega) = \frac{10\pi^5H_s^2}{T_p^4\omega^5}\exp\left(-\frac{20\pi^4}{T_p^4\omega^4}\right),
\end{equation}
where $S(\omega)$ [m$^2$-s] is the wave spectral density as a function of frequency, $\omega$ [rad/s]~\cite{falnes2020}. Two parameters define the sea state: a significant wave height ($H_s$) and peak period ($T_p$). The significant wave height is the average wave height ($2A_{wave}$) of the top one-third of the waves in the sea state, and the peak period denotes the period associated with the maximum spectral energy density. Pierson-Moskowitz wave spectra with 12~m water depth are used for all sea states in the sensitivity study, despite the potential for different spectra shapes and water depths at different locations. While differences in these would be expected at different locations, since this is not a true site selection study and instead just an analysis of the sensitivity of optimal design to sea state this is sufficient.

This formulation reduces the WEC hydrodynamics to four key coefficients: the hydrostatic stiffness ($K_{hs}$), added mass ($A(\omega)$), radiation damping ($B(\omega)$), and excitation force/moment amplitude operator ($F_e$), with the wave spectrum being defined by parameters. These coefficients are determined using Capytaine \cite{ancellin_capytaine_2019}, an open-source boundary element method (BEM) solver. BEM solvers are well-suited for wave-structure interaction analysis because they require meshing of the structure-fluid interface only. Computational fluid dynamics solvers require meshing of the entire fluid domain; this makes BEM solvers significantly less computationally expensive~\cite{bem_wave}, but depends on the validity of the linear potential flow assumptions.

This study approximates the OSWEC as a rectangular prism, with one degree of freedom (pitch about the bottom hinge). Although this is a simplification, similar simplifications are common in other high level early stage WEC design optimization studies. For each design evaluation, a new mesh is created, and the BEM solver finds the hydrodynamic coefficients across 20 frequencies, specified in Table~\ref{tab:paramssolve}. The number of panels used in the mesh is scaled by OSWEC size to ensure enough panels are used to capture the hydrodynamics of larger devices, while not using an excessive number of panels for smaller devices. The equations used to determine the number of panels in the mesh are shown in \ref{app:mesh}.

\subsection{System Dynamics} \label{sec:sysdyn}
The system dynamics module solves the coupled dynamics of multiple disciplines (rigid body dynamics and hydraulics) simultaneously. The interaction between these two subdisciplines is shown in the xDSM in Figure~\ref{fig:xdsm}. These two disciplines are coupled using Simulink, with the piston motion from the rigid body dynamics module driving the hydraulic circuit in the hydraulics module, and the PTO force ($f_u$) from the hydraulics module feeding back into the rigid body dynamics module. MATLAB's ode4 solver, a fourth-order Runge-Kutta solver, with a 0.1~s time step, is used. 

The rigid body dynamics module considers the OSWEC dynamics as well as the dynamics of the mechanism connecting the WEC to the piston. The OSWEC dynamics are governed by Equation~\ref{eq:wec}, with the various hydrodynamic coefficients determined through the BEM solver, as discussed in Section~\ref{sec:hydro}. WEC-Sim~\cite{wecsim}, an open-source software developed by the National Laboratory of the Rockies (NLR) and Sandia National Laboratories, uses these coefficients to solve the WEC dynamics in the time domain, when provided wave spectrum information. The mechanism is modeled using the Simscape multibody toolbox, modeling the kinematics and dynamics of the mechanism. This model relates the OSWEC pitch motion to the linear motion of the piston, as well as the force on the piston to the torque at the OSWEC hinge. 

\begin{figure*}[t]
    \centering
    \includegraphics[width=\linewidth]{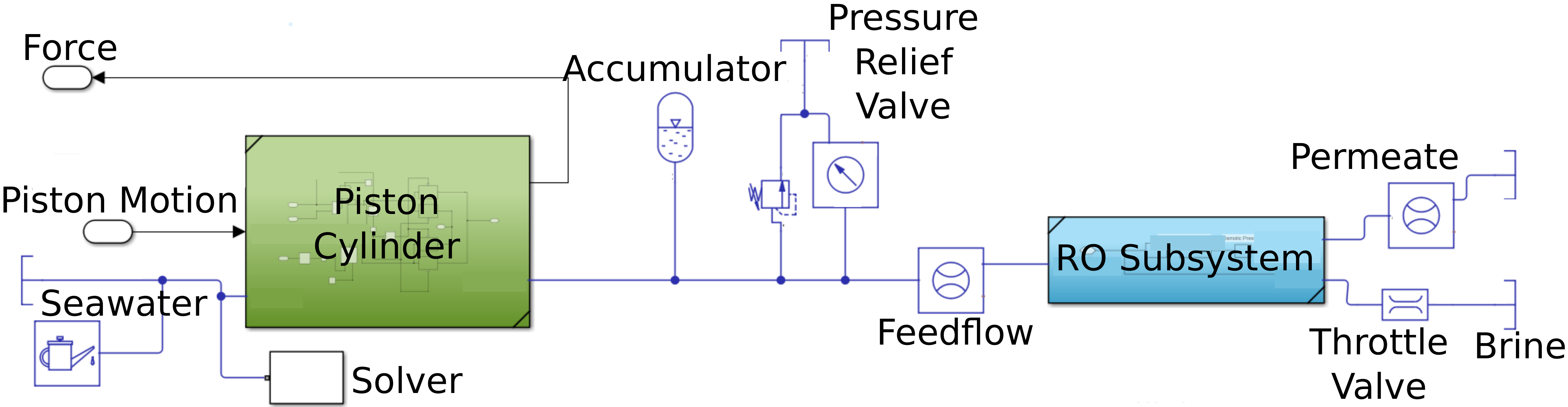}
    \captionof{figure}{Simscape model of the hydraulic circuit.}
    \label{fig:hydraulic_simscape}
\end{figure*}

The hydraulic circuit (Figure~\ref{fig:hydraulics}) is modeled using the isothermal liquid domain from the Simscape fluids toolbox. This domain is well-suited for modeling hydraulic systems. A high-level view of the Simscape model is shown in Figure~\ref{fig:hydraulic_simscape} showing the key components of the hydraulic circuit. The Piston Cylinder models the piston and directional valves. Flow out of the Piston Cylinder then passes through the accumulator, pressure relief valve, the RO Subsystem (which models the SWRO membrane), and finally the throttle valve on the brine side. 
The Piston Cylinder is modeled similarly to the model used by Yu and Jenne (2018)~\cite{Yu2018}. With a single double-acting hydraulic actuator connected to a series of three-way valves. The opening of the three-way valves is dictated by the direction of the piston motion to ensure flow is properly flowing into the cylinder from the ocean and out of the cylinder into the hydraulic circuit and SWRO membrane. Our model replaces the legacy Hydraulic domain used by Yu and Jenne with Simscape’s newer Isothermal Liquid domain, a modernization that improves accuracy and robustness~\cite{ssc_IL}.

\begin{figure}[b]
    \centering
    \includegraphics[width=0.5\linewidth]{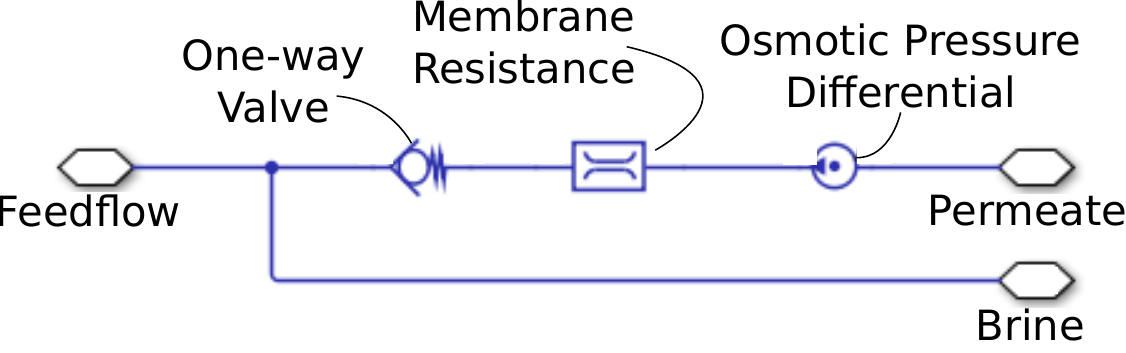}
    \captionof{figure}{Simscape model of the RO membrane.}
    \label{fig:ro_simscape}
\end{figure}

The RO Subsystem model is shown in Figure~\ref{fig:ro_simscape}. The check valve prevents forward osmosis, and the membrane resistance block combined with the osmotic pressure block creates a Simscape representation of Equation \ref{eq:ro}.

A major limitation of the system dynamics model is that it simulates only a single sea state over a 1800~s time period. Although this duration is sufficient to obtain accurate performance metrics for a given sea state, it does not capture longer-term dynamics associated with transitions between sea states over an annual cycle, nor operational procedures such as start-up, shut-down, maintenance, and extended periods of low energy availability. Given the model fidelity and associated computational cost, however, simulating continuous annual operation across varying sea states is impractical within the scope the present optimization study. Future work should therefore either evaluate the annual performance of a small number of candidate designs using a high-fidelity model or employ a lower-fidelity framework that enables long-term operational dynamics to be incorporated directly into the optimization.

\subsection{Economics} \label{sec:econ}
Our objective function is levelized cost of water (LCOW) [\$/m$^3$], an adaptation of the levelized cost of energy (LCOE) [\$/kWh] function used by the U.S. Department of Energy \cite{LCOE_DOE}, 
\begin{equation}
    \text{LCOW} = \frac{(\text{FCR}\times\text{CAPEX}) + \text{OPEX}}{\text{AWP}},
    \label{eq:LCOW}
\end{equation}
\noindent where the fixed charge rate, FCR, is set to 10.8\% as assumed by Alison LaBonte et al. (2013)~\cite{LCOE_DOE}. Annual water production (AWP) [m$^3$/yr] is calculated by scaling the results of the system dynamics simulation to a yearly basis. The costs are then split into capital expenditures (CAPEX) [\$] and annual operating expenses (OPEX) [\$/yr].

To determine both CAPEX and OPEX, expenses are broken down into three cost models: WEC, PTO, and SWRO plant. These cost models are not intended to provide exact cost estimates. Rather, they are used to quantify the relative costs of different designs to guide the optimization process. Although these cost models are not comprehensive, they serve as a tool to highlight the impact of MDO for wave-driven desalination systems. Each cost model is discussed in the following sections. Note that all costs are reported in 2025 USD.

WEC cost modeling is challenging due to limited real-world data, limiting accuracy. Despite this challenge, a practical cost model that captures key trends for design comparison is possible. In this study, we follow the approach used by Grasberger et al. (2024)~\cite{Grasberger2024}. This model defines cost as a function of the float's surface area:
\begin{equation}
    \text{Cost} = C_{1,ref}\left(\frac{A_{x}}{A_{ref}}\right) + C_{2,ref}\left(1 + \text{log}\left(\frac{A_{x}}{A_{ref}}\right)\right),
\end{equation}
\noindent where $A_{x}$ [m$^2$] is the float surface area and $A_{ref}$ [m$^2$] is the reference float surface area. $C_{1,ref}$ [\$] and $C_{2,ref}$ [\$] are the costs of the reference float associated with the two terms. Structural components are captured in the linear term, while other expenses that are less dependent on size are captured in the logarithmic term. This cost model is dependent on accurate reference costs to make meaningful comparisons. For our study, reference costs are taken from NLR's Reference Model 5 (RM5)~\cite{rm5}. The costs from this reference model are tailored to our study by excluding the costs of the electrical PTO system, which is irrelevant in this study. Deployment and installation costs are likewise omitted because the design variables that most strongly affect these costs, including water depth, bathymetry, distance from shore, are held constant as parameters. Consequently, these costs act as a uniform offset across all candidate designs and do not affect the optimization outcome. 

At small WEC sizes (when the float surface area is less than a tenth of the reference area), the logarithmic term becomes negative, which is unrealistic. To avoid this, a minimum $C_2$ cost is used when the equation drops below this threshold. For this study, this minimum $C_2$ cost is set to \$0 for both CAPEX and OPEX.

The PTO cost covers the hydraulic cylinder and accumulator. The accumulator cost is estimated as a function of accumulator volume ($V_{acc}$), given by the following equation (R$^2$ = 0.9998):
\begin{equation}
    \text{Cost} = 1.621 \times 10^5 V_{acc} ^{0.986},
\end{equation}
\noindent where the constants are tuned to approximate supplier quotes \cite{reasontek}. 

The hydraulic cylinder cost is a function of the required amount of steel:
\begin{equation}
    \text{Cost} = C_{\text{316}} \rho_{\text{316}} V_{\text{316}} (1+L),
\end{equation}
\noindent where $C_{\text{316}}$ [\$/lb] is the unit cost of 316 stainless steel, $\rho_{\text{316}}$ [lb/in$^3$] is the density of 316 stainless steel, $V_{\text{316}}$ [in$^3$] is the volume of 316 stainless steel required, and $L$ [-] is a labor multiplier. The volume required is the sum of the required volumes for the various parts of the cylinder: the cylinder barrel ($V_{\text{cylinder}}$) [in$^3$], the end caps ($V_{\text{cap}}$) [in$^3$], the piston head ($V_{\text{piston}}$) [in$^3$], and the piston rod ($V_{\text{rod}}$) [in$^3$]. 
\begin{equation}
    V_{\text{316}} = V_{\text{cylinder}} + 2V_{cap} + V_{\text{piston}} + V_{\text{rod}},
\end{equation}
The thicknesses of the cylinder barrel and end caps are calculated according to the ASME Boiler and Pressure Vessel Code~\cite{ASME_BPVC}. The piston head uses the same thickness as the end caps, while the piston rod diameter is set to avoid buckling and tensile/compressive failure. The required stroke length, which is needed for the cylinder volume calculation, is determined from the system dynamics simulation. The labor multiplier was tuned to match trends from quotes received~\cite{jit_industries}.

SWRO plant cost models on the scale of this paper (producing thousands of m$^3$/day) are scarce and poorly documented. Most well-documented cost models focus on large-scale plants (hundreds of thousands of m$^3$/day). For example, Haefner and Haji estimate costs of SWRO plants ranging from 75,000 to 7,880,000 m$^3$/day~\cite{Haefner2023}. Models intended to be used at these scales are not suitable for this study. Studies that examine plants of comparable size typically lack transparent cost models. Bilton et al. (2011)~\cite{Bilton2011} and Yu and Jenne (2017)~\cite{YJecon2017} both present studies that leverage constant multipliers to estimate the cost based on the plant capacity that are sourced from the Desalination Economic Evaluation Program, but the program itself is not transparent~\cite{DEEP5manual}.

This study uses a transparent cost model for small-scale SWRO plants. Voutchkov’s \emph{Desalination Project Cost Estimating and Management}~\cite{voutch} (see CAPEX in Section 4 and OPEX in Section 5) provides cost curves spanning a wide range of plant capacities. By choosing among different sets of curves, the model can represent different plant architectures, which adds flexibility. In this study, each curve is fitted with a power function to avoid unrealistic behavior at the low-capacity end of the range. Tables containing all the parameters characterizing the cost curves can be found in \ref{app:swro_curves}. These cost curves incorporate costs associated with conventional intake infrastructure and plant operation but do not capture potential deviations attributable to the WEC-driven nature. The omission of intake-related cost adjustments is not expected to affect the optimization results, as the distance from shore is fixed across all candidate designs. However, difference in operating costs associated with WEC-driven desalination may influence the optimal solution and represent an important area for future research~\cite{Sitterley2022}.

\subsection{Model Validation}
\begin{figure}[t!]
    \centering
    \includegraphics[width=\linewidth]{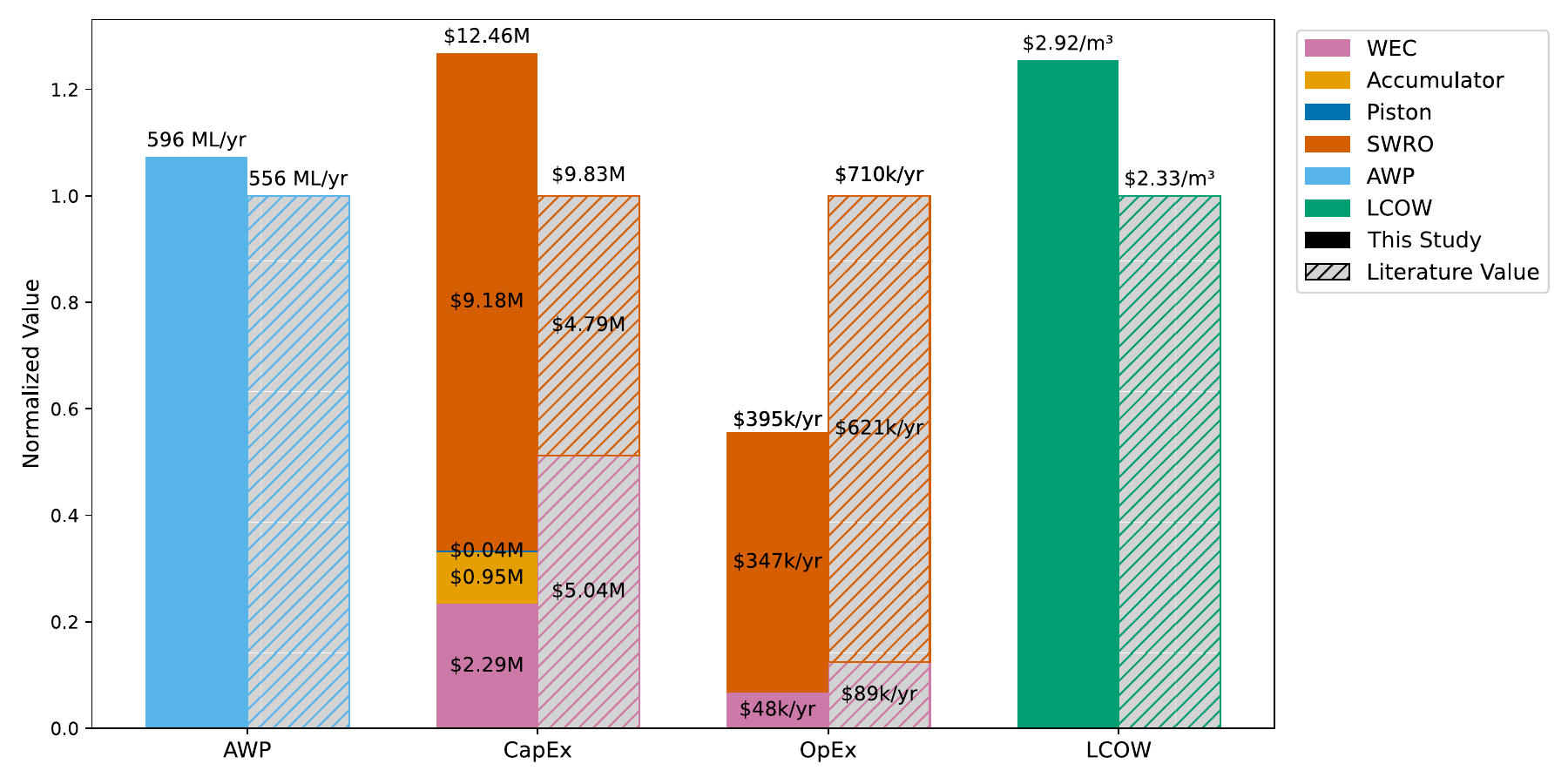}
    \captionof{figure}{Comparison of model results to Yu and Jenne (2017) \cite{YJecon2017}. All costs are reported in 2025 USD.}
    \label{fig:validation}
\end{figure}

Experimental data for wave-driven desalination systems of comparable architecture and scale are currently unavailable. Consequently, validation is limited to comparisons with established techno-economic studies. Our full model is validated against the results of Yu and Jenne (2017), who present a techno-economic analysis of a wave-driven desalination system with an architecture similar to that considered here~\cite{YJecon2017}. The model is evaluated using a comparable design and sea state. To ensure consistency with the assumptions of Yu and Jenne (2017)~\cite{YJecon2017}, who based their cost model on an array of 100 WECs, the OPEX model coefficients ($C_{1,ref}$ and $C_{2,ref}$) are adjusted in the validation study to reflect the same deployment scale. This assumption significantly reduces the OPEX per WEC. The comparison of results is shown in Figure~\ref{fig:validation}. Some notable differences are that our model predicts a significantly higher SWRO plant cost; however, given the level of detail in our SWRO plant cost model, we have confidence in this result. Our model also predicts cheaper WEC costs, this is due to the absence of PTO and deployment/installation costs in our WEC cost model, which are included in the WEC cost in Yu and Jenne (2017)~\cite{YJecon2017}, but given the desalination PTO cost model is separate, and deployment/installation costs are less device dependent, these costs should be omitted for the purposes of this study. 

\section{Results} \label{sec:results}

The results of this study are presented in two parts. First, a case study is presented, optimizing the system design for a single sea state. This case study is used to investigate the impact of MDO by comparing the results to two different SDO approaches.  Next, the results of the sensitivity analysis are presented, showing how the optimal design changes in different sea states.

\subsection{Case Study: MDO vs SDO}
For our case study, optimization is performed under a single sea state defined by a Pierson-Moskowitz spectrum with a significant wave height of 2.64 m and a peak period of 9.86 s, matching one of the conditions used by Yu and Jenne (2018)~\cite{Yu2018}. The MDO optimization is initialized using the design reported by DeGoede and Haji (2025)~\cite{idetc2025}. While this design was previously identified as optimal, it exhibited premature convergence behavior, motivating its use here as a non-ideal starting point. The optimization converges after 345 generations, yielding a design with an LCOW of \$1.21/m$^3$, a 69.5\% improvement over the nominal design from literature~\cite{YJecon2017} under the assumptions adopted in the present modeling framework. The corresponding optimal design vector is provided in Table~\ref{tab:opt}.
\begin{table}[t!]
    \scriptsize
    \centering
    \caption{Optimized design vector compared to nominal design from literature \cite{YJecon2017}, the optimal design from DeGoede and Haji (2025) \cite{idetc2025} (which was used as the initial design in this optimization), and the SDO results.}
    \begin{tabular}{cccccc}
        \hline
        \textbf{Variable} & \textbf{Nominal} \cite{YJecon2017} & \textbf{SDO A} & \textbf{SDO B} & \textbf{Initial} \cite{idetc2025} & \textbf{MDO} \\
        \hline
        $w$ & 18 m & 4.0 m & 4.0 m & 11.3 m & 4.07 m \\
        $t$ & 1.8 m & 2.51 m & 2.51 m & 1.99 m & 0.83 m \\
        $m$ & 127$\times10^3$ kg & 71$\times10^3$ kg & 71$\times10^3$ kg & 396$\times10^3$ kg & 219$\times10^3$ kg \\
        $\ell_1$ & 1.9 m & 1.38 m & 3.83 m & 3.25 m & 2.68 m \\
        $A_p$ & 0.26 m$^2$ & 0.855 m$^2$ & 0.404 m$^2$ & 0.859 m$^2$ & 0.746 m$^2$ \\
        $V_{acc}$ & 6 m$^3$ & 0.29 m$^3$ & 2.41 m$^3$ & 4.57 m$^3$ & 2.45 m$^3$ \\
        $P_0$ & 3.00 MPa & 5.92 MPa & 5.91 MPa & 5.95 MPa & 5.73 MPa \\
        $Q_{p,max}$ & 3100 m$^3$/day & 6753 m$^3$/day & 1000 m$^3$/day & 4882 m$^3$/day & 6612 m$^3$/day \\
        \hline
        \textbf{LCOW} & \$3.97/m$^3$ & \$2.39/m$^3$ & \$2.17/m$^3$ & \$1.68/m$^3$ & \$1.21/m$^3$ \\
        \hline
    \end{tabular}
    \label{tab:opt}
\end{table}

\begin{figure}[t!]
\centering
\includegraphics[width=\linewidth]{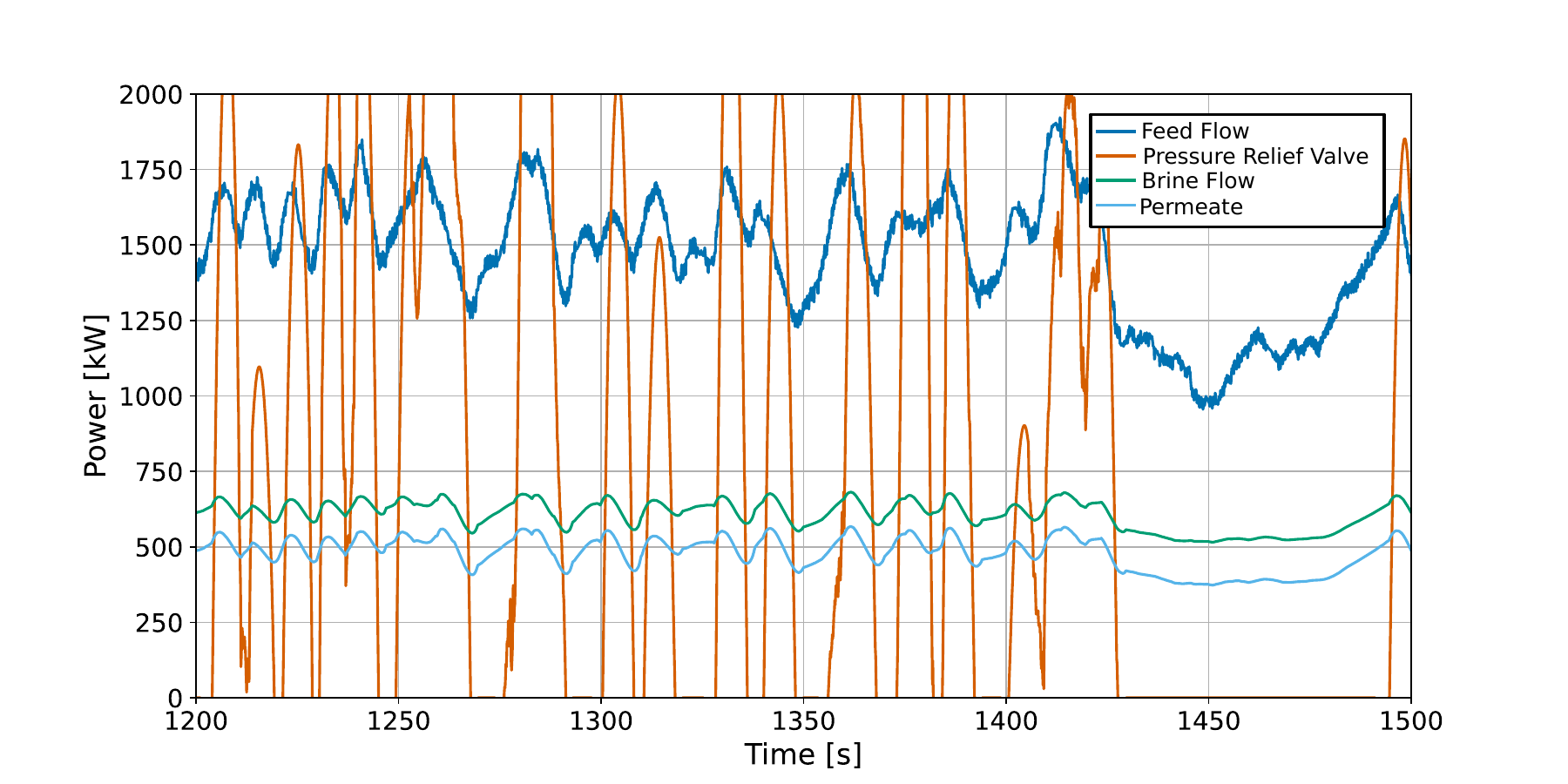}
\captionof{figure}{Power time series of feed flow, pressure relief valve, brine flow, and permeate stream for the MDO-generated design in the case study.}
\label{fig:power}
\end{figure}

\begin{figure}[t]
\centering
\includegraphics[width=0.4\linewidth]{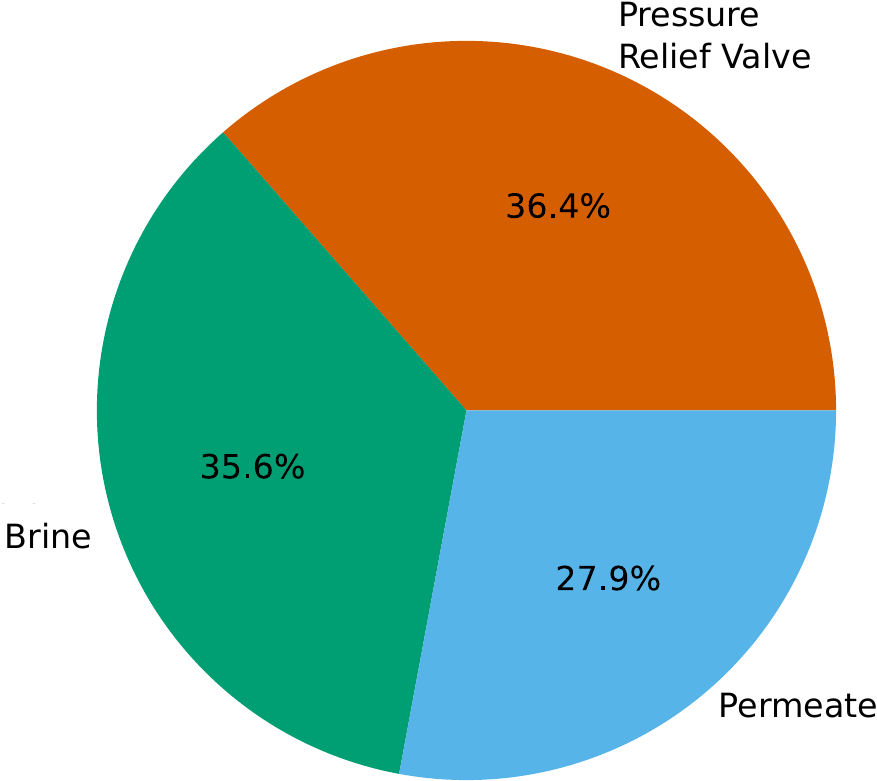}
\captionof{figure}{Breakdown of feed-flow energy in the SWRTO PTO showing the fraction of energy discharged through the pressure relief valve, dissipated in the brine-side throttling valve, or transferred through the SWRO membrane.}
\label{fig:energy}
\end{figure}

The power flowing through different sections of the SWRO PTO is shown in Figure \ref{fig:power}. The figure shows the pressure relief valve dissipating energy when the capacity of the SWRO plant is exceeded, as well as an extended period during which the system operates at lower power and the pressure relief valve remains closed. An integrated energy pie chart is shown in Figure \ref{fig:energy}. This chart more clearly illustrates the distribution of energy usage and highlights two major areas, pressure relief valve and brine flow, where energy efficiency could be increased through an alternative PTO design. Notably these plots show that 36.4\% of the energy entering the system as pressurized seawater leaves the system before reaching the SWRO membranes. A naive response to this result would be to increase the accumulator volume as this would smooth out some of the high energy flow pulses increasing the fraction of seawater energy reaching the membrane. However, the optimizer suggests that this design featuring a smaller accumulator is preferred, meaning the smaller accumulator improves the match between WEC and PTO.

The MDO results suggest some major design changes compared to the nominal design from literature~\cite{YJecon2017}. The optimal design has a 77\% smaller WEC width (4.07 m vs 18 m), a 54\% smaller WEC thickness (0.83 m vs 1.8 m), and a 72\% heavier WEC mass (219$\times10^3$ kg vs 127$\times10^3$ kg). The optimal design also has a 187\% larger piston area (0.746 m$^2$ vs 0.26 m$^2$) and a 41\% larger $\ell_1$ link length (2.68 m vs 1.9 m), a 59\% smaller accumulator volume (2.45 m$^3$ vs 6 m$^3$), a 91\% larger accumulator precharge pressure (5.73 MPa vs 3.00 MPa), and a 113\% larger SWRO plant capacity (6612 m$^3$/day vs 3100 m$^3$/day).

\begin{figure}[b!]
    \centering
    \includegraphics[width=0.9\linewidth]{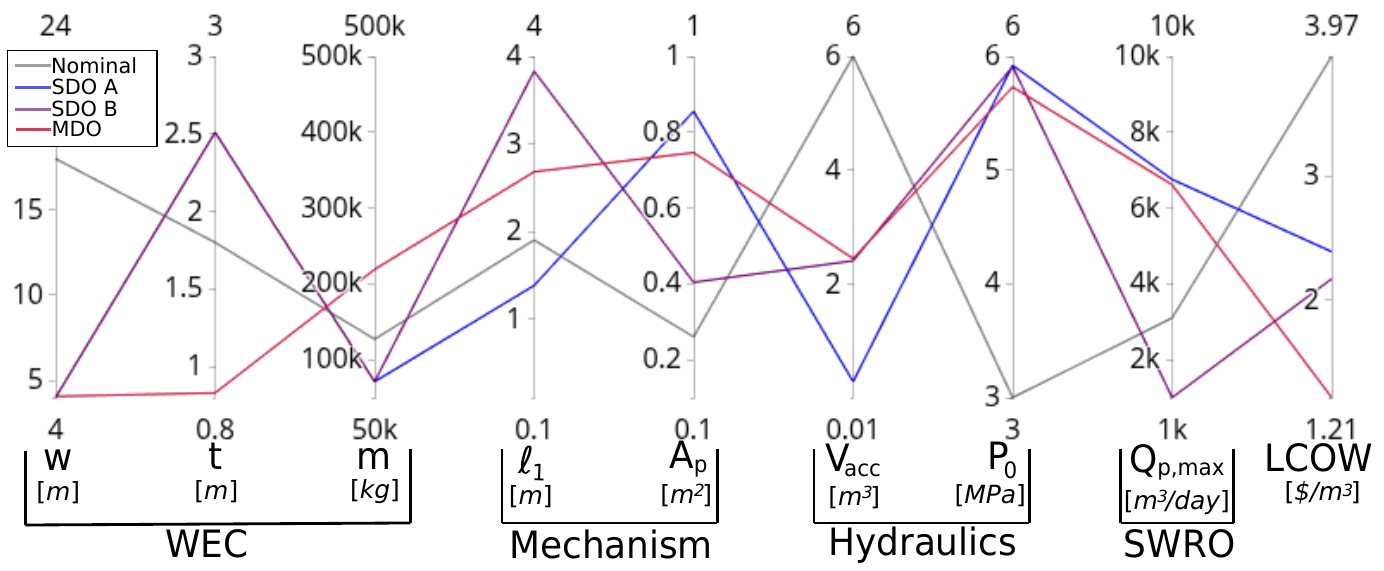}
    \captionof{figure}{Parallel axis plot showing the optimal design variables and LCOW for the MDO and SDO approaches compared to the nominal design from literature \cite{YJecon2017}.}
    \label{fig:par_axis}
\end{figure}

For comparison, the SDO cases are also summarized in Table~\ref{tab:opt}. Both SDO approaches are given the nominal design reported in the literature~\cite{YJecon2017} as an initial design. This choice ensures a consistent baseline, as the design from DeGoede and Haji (2025)~\cite{idetc2025} (the initial design in the MDO optimization) was obtained via MDO starting from the same nominal configuration. These results are represented graphically in Figure~\ref{fig:par_axis}.

The SDO results demonstrate some patterns. The WECs optimized for LCOKE in isolation also have small widths, but 202\% larger thicknesses (2.51 m vs 0.83 m) and 68\% smaller masses (71$\times10^3$ kg vs 219$\times10^3$ kg) than the MDO design. Depending on the SDO approach, the suggested design for the PTO and SWRO plant is different. In SDO approach A, the optimal design has an 88\% smaller accumulator volume (0.29 m$^3$ vs 2.45 m$^3$) despite a similarly sized SWRO plant (6753 m$^3$/day vs 6612 m$^3$/day) compared to the MDO design. In SDO approach B, the optimal design has a 46\% smaller piston area (0.404 m$^2$ vs 0.746 m$^2$) and an 85\% smaller SWRO plant capacity (1000 m$^3$/day vs 6612 m$^3$/day) than the MDO design. Additionally, while the SDO approaches did find designs with improved LCOW compared to the nominal design from literature~\cite{YJecon2017}, they were not as effective as the MDO approach, with LCOWs of \$2.39/m$^3$ and \$2.17/m$^3$ for SDO approaches A and B, respectively, compared to \$1.21/m$^3$ for the MDO design.

\subsection{Sensitivity Analysis Results}
\begin{figure}[b!]
    \centering
    \includegraphics[width=0.8\linewidth, trim=0cm 1.5cm 0cm 0cm, clip]{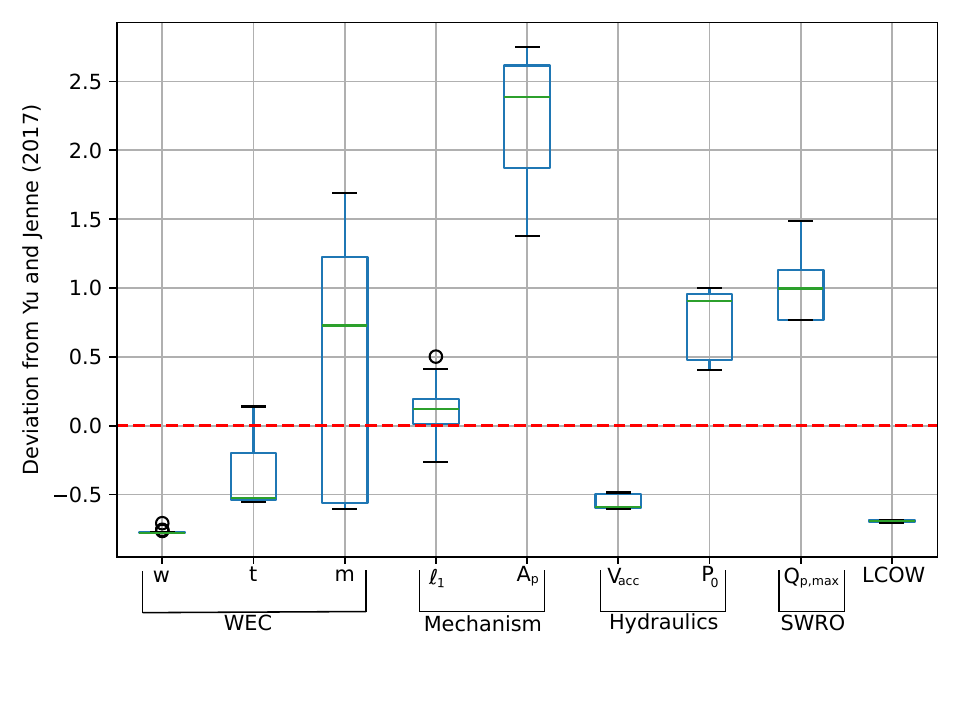}
    \captionof{figure}{Box and whisker plot showing the distribution of optimal design variables and LCOW across 20 different sea states. Green lines represent the mean, blue boxes represent the quartiles, black lines represent the non-outlier limits, and circles represent the outliers.}
    \label{fig:sensitivity}
\end{figure}
The MDO results for a variety of sea states are shown in Figure~\ref{fig:sensitivity}. Each optimization was given the optimal design from DeGoede and Haji (2025)~\cite{idetc2025} as an initial design. The optimal system designs always have smaller WEC widths, larger piston areas, smaller accumulators, and larger SWRO plant capacities, when compared to the nominal design. Additionally, the low variation in LCOW across sea states suggests that although each sea state produces a distinct optimal design, the resulting trade-offs in system configuration converge to a similar overall cost of water.

\section{Discussion} \label{sec:discussion}
The first notable result is the large reduction in LCOW achieved through MDO. The optimized design reaches \$1.21/m$^3$, a 69.5\% improvement over the nominal design from literature~\cite{YJecon2017} under the assumptions adopted in the present modeling framework, supporting the MDO framework as an effective framework for wave-driven desalination system design. 

The case study also shows that MDO significantly outperforms the two SDO approaches for this application. The SDO approaches optimize the WEC in isolation, maximizing kinetic energy capture without accounting for PTO and SWRO interactions. While WEC widths are similar, thickness and mass differ substantially, leading to very different WEC impedance characteristics. The thicker (2.51 m vs 0.83 m) and lighter (71$\times10^3$ kg vs 219$\times10^3$ kg) WECs from SDO improve impedance matching in isolation but perform poorly when coupled with the PTO and SWRO system.

The two SDO approaches also produce different PTO and SWRO designs. In SDO approach A, a small accumulator (0.29 m$^3$) preserves the isolated WEC impedance but requires a large SWRO plant (6753 m$^3$/day) to avoid wasting feedwater, resulting in a high LCOW (\$2.39/m$^3$), nearly twice as high as the MDO design (\$1.21/m$^3$). In SDO approach B, a large accumulator (6 m$^3$) reduces energy capture and flow variability, driving the SWRO plant to its minimum allowed (1000 m$^3$/day) and requiring a smaller piston area (0.404 m$^2$). Despite a smaller SWRO plant capacity and improved isolated WEC performance, this design still yields a higher LCOW (\$2.17/m$^3$) than the MDO design (\$1.21/m$^3$). 

Compared to the nominal design~\cite{YJecon2017}, the results suggest notable design shifts, though these must be interpreted cautiously without fully accounting for sea state variability. Trends across sea states indicate smaller WEC widths and larger piston areas, suggesting the nominal design suffers from a torque mismatch, with the wave excitation exceeding PTO torque. Therefore, smaller WECs (which experience lower wave excitation torques) paired with larger piston areas (which produce higher PTO torques) are more effective. This aligns with the nominal design's origin in the RM5~\cite{rm5}, which  targets much higher pressures (28 MPa) than required by the SWRO membrane only demands ($\sim$5 MPa). 


A second trend in the MDO results is the pairing of smaller accumulators with larger SWRO plant capacities, revealing an important trade-off between flow smoothing and freshwater production. Larger accumulators have traditionally been favored because they reduce flow variability delivered to the SWRO plant and allow excess flow to be stored rather than lost when the plant reaches its fixed capacity. Because SWRO plant capacity is a major cost driver~\cite{YJecon2017}, this smoothing can make larger accumulators appear economically attractive by reducing the required plant size and increasing capacity utilization. This perspective, however, considers only the desalination system and neglects the influence of accumulator size on PTO dynamics. In the coupled WEC-PTO-SWRO system, accumulator size affects PTO impedance and therefore alters WEC energy capture. The MDO results indicate that smaller accumulators improve impedance matching and increase absorbed wave energy, despite requiring larger SWRO plant capacities to accommodate the resulting flow variability as well as increased energy losses through the pressure relief valve. Figure \ref{fig:energy} shows that 36.4\% of the energy captured by the WEC PTO is lost to the pressure relief valve and while these losses could easily be reduced by increasing the accumulator size, the optimizer finds this seemingly inefficient design to be better. The optimized designs therefore suggest that the benefits of increased energy capture outweigh the additional cost of larger SWRO plants and energy lost to the pressure relief valve. More broadly, these findings highlight the importance of accounting for system-level interactions and suggest that designs with less aggressive flow smoothing may be preferable in integrated wave-driven desalination systems.

These trends are only revealed through MDO, which captures the coupled dynamics of the WEC, PTO, and SWRO plant. This underscores the importance of integrated MDO framework for designing wave-driven desalination systems, as key tradeoffs and system-level optima emerge only when all subsystems are considered together. Of course, the actual annual performance of these devices is likely to differ from the model's predictions, particularly due to the impact variations in sea states over time will have on SWRO plant operational costs and annual water production.

\section{Conclusion} \label{sec:conclusion}

This study presents an MDO framework for designing wave-driven desalination systems. For the representative case study, MDO reduced LCOW from \$3.97/m$^3$ for the nominal design to \$1.21/m$^3$, corresponding to a 69.5\% reduction within the assumptions of the present modeling framework. The MDO solution also substantially outperformed the two SDO approaches, which yielded LCOW values of \$2.39/m$^3$ and \$2.17/m$^3$, respectively. Sensitivity analysis across 20 sea states from five locations further showed that MDO consistently recommends key design shifts within this modeling framework including smaller WECs, larger piston areas, smaller accumulators, and increased SWRO plant capacity. These trends emerge only when using a holistic, coupled-system model underscoring the value of MDO for this application.

This study is not without limitations. Although Sitterley et al. (2022) suggest transient SWRO operation is feasible~\cite{Sitterley2022}, uncertainties remain regarding long-term membrane performance and lifespan. Additionally, the cost models used in this study are intended to capture key trends for design comparison rather than provide precise estimates. It would be valuable to investigate the sensitivity of optimal designs to economic parameters, in order to quantify how robust the observed design trends are to uncertainty in the cost model. It would also be valuable to investigate how alternative PTO designs affect these results. In the case study, the MDO suggested design saw only 27.9\% of the input energy reach the permeate stream, suggesting significant opportunities for efficiency improvements through the incorporation of energy recovery devices on the pressure relief valve discharge and brine flows. Temporal variability in sea states over an annual cycle is also neglected; future work should therefore consider either non-optimizing annual performance assessments, or lower-fidelity models that can capture these dynamics, as they are expected to influence both SWRO plant operation and AWP. Despite these limitations, the framework offers a strong foundation for improving wave-driven desalination system design. Further advances in control strategies, transient desalination modeling, and physical implementation will be key to accelerating the development and deployment of these systems as a sustainable response to the global water crisis. 

\section*{Data Availability}
The code and results for this manuscript are open source and available at \url{https://github.com/symbiotic-engineering/mdo_wd2/releases/tag/RE2026}.

\section*{CRediT authorship contribution statement}
\textbf{Nate DeGoede}: Conceptualization, Formal analysis, Investigation, Methodology, Software, Validation, Writing – original draft, Writing – review \& editing. \textbf{Maha Haji}: Funding acquisition, Project administration, Resources, Supervision, Writing – review \& editing.

\section*{Declaration of Competing Interest}
The authors declare that they have no known competing financial interests or personal relationships that could have appeared to influence the work reported in this paper.

\section*{Acknowledgements}
The authors would like to thank MathWorks for providing funding support and technical support for this project. The authors would also like to thank Georgia McSwain and Olivia Vitale for their valuable feedback on the manuscript.

\bibliographystyle{elsarticle-num}

\begin{thebibliography}{10}
\expandafter\ifx\csname url\endcsname\relax
  \def\url#1{\texttt{#1}}\fi
\expandafter\ifx\csname urlprefix\endcsname\relax\def\urlprefix{URL }\fi
\expandafter\ifx\csname href\endcsname\relax
  \def\href#1#2{#2} \def\path#1{#1}\fi

\bibitem{watershortage2015}
J.~Eliasson, The rising pressure of global water shortages, Nature 517~(6)
  (Jan. 2015).
\newblock \href {https://doi.org/https://doi.org/10.1038/517006a}
  {\path{doi:https://doi.org/10.1038/517006a}}.

\bibitem{Li2018}
Z.~Li, A.~Siddiqi, L.~D. Anadon, V.~Narayanamurti, Towards sustainability in
  water-energy nexus: Ocean energy for seawater desalination, Renewable and
  Sustainable Energy Reviews 82 (2018) 3833--3847.
\newblock \href {https://doi.org/10.1016/j.rser.2017.10.087}
  {\path{doi:10.1016/j.rser.2017.10.087}}.

\bibitem{blue_econ}
{U.S. Department of Energy}, Powering the blue economy: Exploring opportunities
  for marine renewable energy in maritime markets (4 2019).

\bibitem{nytdrought}
J.~Gillis,
  \href{http://web-static-aws.seas.harvard.edu/climate/eli/Courses/global-change-debates/Sources/California-droughts/California
  Drought Is Made Worse by Global Warming, Scientists Say - The New York
  Times.pdf}{California drought is made worseby global warming, scientists
  say}, New York Times (Aug. 2015).
\newline\urlprefix\url{http://web-static-aws.seas.harvard.edu/climate/eli/Courses/global-change-debates/Sources/California-droughts/California
  Drought Is Made Worse by Global Warming, Scientists Say - The New York
  Times.pdf}

\bibitem{Lai2016}
W.~Lai, Q.~Ma, H.~Lu, S.~Weng, J.~Fan, H.~Fang, Effects of wind intermittence
  and fluctuation on reverse osmosis desalination process and solution
  strategies, Desalination 395 (2016) 17--27.
\newblock \href {https://doi.org/10.1016/j.desal.2016.05.019}
  {\path{doi:10.1016/j.desal.2016.05.019}}.

\bibitem{Elkadeem2024}
M.~R. Elkadeem, K.~M. Kotb, S.~W. Sharshir, M.~A. Hamada, A.~E. Kabeel, I.~K.
  Gabr, M.~A. Hassan, M.~Y. Worku, M.~A. Abido, Z.~Ullah, H.~M. Hasanien, F.~F.
  Selim, Optimize and analyze a large-scale grid-tied solar pv-powered swro
  system for sustainable water-energy nexus, Desalination 579 (2024) 117440.
\newblock \href {https://doi.org/10.1016/j.desal.2024.117440}
  {\path{doi:10.1016/j.desal.2024.117440}}.

\bibitem{Davies2005}
P.~Davies, Wave-powered desalination: resource assessment and review of
  technology, Desalination 186~(1–3) (2005) 97--109.
\newblock \href {https://doi.org/10.1016/j.desal.2005.03.093}
  {\path{doi:10.1016/j.desal.2005.03.093}}.

\bibitem{Falcao2010}
A.~F. d.~O. Falcão, Wave energy utilization: A review of the technologies,
  Renewable and Sustainable Energy Reviews 14~(3) (2010) 899--918.
\newblock \href {https://doi.org/10.1016/j.rser.2009.11.003}
  {\path{doi:10.1016/j.rser.2009.11.003}}.

\bibitem{Sobieski}
J.~Sobieszczanski-Sobieski, Multidisciplinary design optimization: An emerging
  new engineering discipline, Solid Mechanics and Its Applications (1995)
  483–496\href {https://doi.org/10.1007/978-94-011-0453-1_14}
  {\path{doi:10.1007/978-94-011-0453-1_14}}.

\bibitem{PenaSanchez2022}
Y.~Peña-Sanchez, D.~García-Violini, J.~V. Ringwood, Control co-design of
  power take-off parameters for wave energy systems, IFAC-PapersOnLine 55~(27)
  (2022) 311--316.
\newblock \href {https://doi.org/10.1016/j.ifacol.2022.10.531}
  {\path{doi:10.1016/j.ifacol.2022.10.531}}.

\bibitem{Rosati2023}
M.~Rosati, J.~V. Ringwood, Control co-design of power take-off and bypass valve
  for owc-based wave energy conversion systems, Renewable Energy 219 (2023)
  119523.
\newblock \href {https://doi.org/10.1016/j.renene.2023.119523}
  {\path{doi:10.1016/j.renene.2023.119523}}.

\bibitem{Stroefer2023}
C.~A. Michelén-Ströfer, D.~T. Gaebele, R.~G. Coe, G.~Bacelli, Control
  co-design of power take-off systems for wave energy converters using
  wecopttool, IEEE Transactions on Sustainable Energy 14~(4) (2023) 2157--2167.
\newblock \href {https://doi.org/10.1109/tste.2023.3272868}
  {\path{doi:10.1109/tste.2023.3272868}}.

\bibitem{Grasberger2024}
J.~Grasberger, L.~Yang, G.~Bacelli, L.~Zuo, Control co-design and optimization
  of oscillating-surge wave energy converter, Renewable Energy 225 (2024)
  120234.
\newblock \href {https://doi.org/10.1016/j.renene.2024.120234}
  {\path{doi:10.1016/j.renene.2024.120234}}.

\bibitem{GarciaRosa2016}
P.~B. Garcia-Rosa, J.~V. Ringwood, On the sensitivity of optimal wave energy
  device geometry to the energy maximizing control system, IEEE Transactions on
  Sustainable Energy 7~(1) (2016) 419--426.
\newblock \href {https://doi.org/10.1109/TSTE.2015.2423551}
  {\path{doi:10.1109/TSTE.2015.2423551}}.

\bibitem{Coe2021}
R.~G. Coe, G.~Bacelli, D.~Forbush, A practical approach to wave energy modeling
  and control, Renewable and Sustainable Energy Reviews 142~(represent) (2021)
  110791.
\newblock \href {https://doi.org/10.1016/j.rser.2021.110791}
  {\path{doi:10.1016/j.rser.2021.110791}}.

\bibitem{Coe2025}
R.~G. Coe, G.~Bacelli, D.~Gaebele, A.~Keow, D.~Forbush,
  \href{https://www.sciencedirect.com/science/article/pii/S0957415825001047}{Co-design
  of a wave energy converter through bi-conjugate impedance matching},
  Mechatronics 111 (2025) 103395.
\newblock \href
  {https://doi.org/https://doi.org/10.1016/j.mechatronics.2025.103395}
  {\path{doi:https://doi.org/10.1016/j.mechatronics.2025.103395}}.
\newline\urlprefix\url{https://www.sciencedirect.com/science/article/pii/S0957415825001047}

\bibitem{Ladan2015}
S.~Ladan, K.~Wu, Nonlinear modeling and harmonic recycling of millimeter-wave
  rectifier circuit, IEEE Transactions on Microwave Theory and Techniques
  63~(3) (2015) 937--944.
\newblock \href {https://doi.org/10.1109/TMTT.2015.2396043}
  {\path{doi:10.1109/TMTT.2015.2396043}}.

\bibitem{Ylaenen2014}
M.~M. Ylänen, M.~J. Lampinen, Determining optimal operating pressure for
  aaltoro – a novel wave powered desalination system, Renewable Energy 69
  (2014) 386--392.
\newblock \href {https://doi.org/10.1016/j.renene.2014.03.061}
  {\path{doi:10.1016/j.renene.2014.03.061}}.

\bibitem{YJecon2017}
Y.-H. Yu, D.~Jenne, Analysis of a wave-powered, reverse-osmosis system and its
  economic availability in the united states, in: Proceedings of the ASME 2017
  36th International Conference on Ocean, Offshore and Arctic Engineering
  OMAE2017, 2017.
\newblock \href {https://doi.org/https://doi.org/10.1115/OMAE2017-62136}
  {\path{doi:https://doi.org/10.1115/OMAE2017-62136}}.

\bibitem{Yu2018}
Y.-H. Yu, D.~Jenne, Numerical modeling and dynamic analysis of a wave-powered
  reverse-osmosis system, Journal of Marine Science and Engineering 6~(4)
  (2018) 132.
\newblock \href {https://doi.org/10.3390/jmse6040132}
  {\path{doi:10.3390/jmse6040132}}.

\bibitem{brodersen2022}
K.~M. Brodersen, E.~A. Bywater, A.~M. Lanter, H.~H. Schennum, K.~N. Furia,
  M.~K. Sheth, N.~S. Kiefer, B.~K. Cafferty, A.~K. Rao, J.~M. Garcia, D.~M.
  Warsinger, Direct-drive ocean wave-powered batch reverse osmosis,
  Desalination 523 (2022) 115393.
\newblock \href {https://doi.org/10.1016/j.desal.2021.115393}
  {\path{doi:10.1016/j.desal.2021.115393}}.

\bibitem{Suchithra2022}
R.~Suchithra, T.~K. Das, K.~Rajagopalan, A.~Chaudhuri, N.~Ulm, M.~Prabu,
  A.~Samad, P.~Cross, Numerical modelling and design of a small-scale
  wave-powered desalination system, Ocean Engineering 256 (2022) 111419.
\newblock \href {https://doi.org/10.1016/j.oceaneng.2022.111419}
  {\path{doi:10.1016/j.oceaneng.2022.111419}}.

\bibitem{Simmons2023}
J.~W. Simmons, J.~D. Van~de Ven,
  \href{https://www.mdpi.com/1996-1073/16/21/7381}{A comparison of power
  take-off architectures for wave-powered reverse osmosis desalination of
  seawater with co-production of electricity}, Energies 16~(21) (2023).
\newblock \href {https://doi.org/10.3390/en16217381}
  {\path{doi:10.3390/en16217381}}.
\newline\urlprefix\url{https://www.mdpi.com/1996-1073/16/21/7381}

\bibitem{Robertson2022}
B.~Robertson, G.~Dunkle, T.~Mundon, L.~Kilcher,
  \href{https://marineenergyjournal.org/imej/article/view/104}{Wave resource
  spatial and temporal variability dependence on wec size}, International
  Marine Energy Journal 5~(1) (2022) 113–121.
\newblock \href {https://doi.org/10.36688/imej.5.113-121}
  {\path{doi:10.36688/imej.5.113-121}}.
\newline\urlprefix\url{https://marineenergyjournal.org/imej/article/view/104}

\bibitem{Mi2023}
J.~Mi, X.~Wu, J.~Capper, X.~Li, A.~Shalaby, R.~Wang, S.~Lin, M.~Hajj, L.~Zuo,
  Experimental investigation of a reverse osmosis desalination system directly
  powered by wave energy, Applied Energy 343 (2023) 121194.
\newblock \href {https://doi.org/10.1016/j.apenergy.2023.121194}
  {\path{doi:10.1016/j.apenergy.2023.121194}}.

\bibitem{xdsm}
A.~B. Lambe, J.~R. R.~A. Martins,
  \href{https://doi.org/10.1007/s00158-012-0763-y}{Extensions to the design
  structure matrix for the description of multidisciplinary design, analysis,
  and optimization processes}, Structural and Multidisciplinary Optimization
  46~(2) (2012) 273--284.
\newblock \href {https://doi.org/10.1007/s00158-012-0763-y}
  {\path{doi:10.1007/s00158-012-0763-y}}.
\newline\urlprefix\url{https://doi.org/10.1007/s00158-012-0763-y}

\bibitem{Arthur2007}
D.~Arthur, S.~Vassilvitskii, k-means++: the advantages of careful seeding, in:
  Proceedings of the Eighteenth Annual ACM-SIAM Symposium on Discrete
  Algorithms, SODA '07, Society for Industrial and Applied Mathematics, USA,
  2007, p. 1027–1035.

\bibitem{ndbc}
{U.S. National Data Buoy Center}, National data buoy center – real-time
  oceanographic and meteorological data, \url{https://www.ndbc.noaa.gov/}
  (1971).

\bibitem{idetc2025}
N.~DeGoede, M.~Haji, \href{https://doi.org/10.1115/DETC2025-168312}{A
  multidisciplinary design optimization framework for wave-driven desalination
  systems}, Vol. Volume 3B: 51st Design Automation Conference (DAC) of
  International Design Engineering Technical Conferences and Computers and
  Information in Engineering Conference, 2025, p. V03BT03A010.
\newblock \href
  {http://arxiv.org/abs/https://asmedigitalcollection.asme.org/IDETC-CIE/proceedings-pdf/IDETC-CIE2025/89237/V03BT03A010/7557602/v03bt03a010-detc2025-168312.pdf}
  {\path{arXiv:https://asmedigitalcollection.asme.org/IDETC-CIE/proceedings-pdf/IDETC-CIE2025/89237/V03BT03A010/7557602/v03bt03a010-detc2025-168312.pdf}},
  \href {https://doi.org/10.1115/DETC2025-168312}
  {\path{doi:10.1115/DETC2025-168312}}.
\newline\urlprefix\url{https://doi.org/10.1115/DETC2025-168312}

\bibitem{separationprocesses}
J.~D. Seader, E.~J. Henley, D.~K. Roper, SEPARATION PROCESS PRINCIPLES:
  Chemical and Biochemical Operations, 3rd Edition, Wiley, 2011.

\bibitem{Kurby2025}
K.~A. Sitterley, Z.~Binger, D.~S. Jenne, Comparing constant and transient
  membrane transport parameters for use in wave desalination models, 2025.
\newblock \href {https://doi.org/.org/10.3390/membranes15080243}
  {\path{doi:.org/10.3390/membranes15080243}}.

\bibitem{SW30HR380}
DuPont, Filmtec™ sw30hr-380 element product data sheet, Datasheet (Oct.
  2024).

\bibitem{wave}
DuPont, Wave design software.

\bibitem{Bacelli2009}
G.~Bacelli, J.-C. Gilloteaux, J.~Ringwood, A predictive controller for a
  heaving buoy producing potable water, in: European Control Conference, 2009.

\bibitem{Sitterley2022}
K.~A. Sitterley, T.~J. Cath, D.~S. Jenne, Y.-H. Yu, T.~Y. Cath, Performance of
  reverse osmosis membrane with large feed pressure fluctuations from a
  wave-driven desalination system, Desalination 527 (2022) 115546.
\newblock \href {https://doi.org/10.1016/j.desal.2022.115546}
  {\path{doi:10.1016/j.desal.2022.115546}}.

\bibitem{falnes2020}
J.~Falnes, A.~Kurniawan, Ocean Waves and Oscillating Systems: Linear
  Interactions Including Wave-Energy Extraction, 2nd Edition, Cambridge
  University Press, 2020.
\newblock \href {https://doi.org/10.1017/9781108674812}
  {\path{doi:10.1017/9781108674812}}.

\bibitem{wecsim}
\href{https://wec-sim.github.io/WEC-Sim/master/index.html}{Wec-sim (wave energy
  converter simulator)}.
\newline\urlprefix\url{https://wec-sim.github.io/WEC-Sim/master/index.html}

\bibitem{ancellin_capytaine_2019}
M.~Ancellin, F.~Dias, \href{https://doi.org/10.21105%2Fjoss.01341}{Capytaine: a
  {Python-based} linear potential flow solver}, Journal of Open Source Software
  4~(36) (2019) 1341.
\newblock \href {https://doi.org/10.21105/joss.01341}
  {\path{doi:10.21105/joss.01341}}.
\newline\urlprefix\url{https://doi.org/10.21105%2Fjoss.01341}

\bibitem{bem_wave}
J.~Newman, C.-H. Lee, Boundary-element methods in offshore structure analysis,
  Journal of Offshore Mechanics and Arctic Engineering 124 (2002) 81.
\newblock \href {https://doi.org/10.1115/1.1464561}
  {\path{doi:10.1115/1.1464561}}.

\bibitem{ssc_IL}
{MathWorks},
  \href{https://www.mathworks.com/help/simscape/ug/advantages-of-using-isothermal-liquid-blocks.html}{Advantages
  of using isothermal liquid blocks}, accessed: 2026-04-11 (2025).
\newline\urlprefix\url{https://www.mathworks.com/help/simscape/ug/advantages-of-using-isothermal-liquid-blocks.html}

\bibitem{LCOE_DOE}
A.~LaBonte, P.~O’Connor, C.~Fitzpatrick, K.~Hallett, Y.~Li, Standardized cost
  and performance reporting for marineand hydrokinetic technologies, in: 1st
  Marine Energy Technology Symposiumt, Washington DC, 2013.

\bibitem{rm5}
Y.-H. Yu, D.~Jenne, R.~Thresher, A.~Copping, S.~Geerlofs, L.~Hanna, Reference
  model 5 (rm5): Oscillating surge wave energy converter, Tech. Rep. Report
  (Jan. 2015).

\bibitem{reasontek}
ReasonTek, Low pressure accuumulator quote (03 2025).

\bibitem{ASME_BPVC}
{American Society of Mechanical Engineers}, Boiler and Pressure Vessel Code,
  Section VIII, Division 1: Rules for Construction of Pressure Vessels, ASME,
  New York, 2015, aSME BPVC-VIII-1-2015.

\bibitem{jit_industries}
Jit industries: Hydraulic cylinder quotes (2025).

\bibitem{Haefner2023}
M.~W. Haefner, M.~N. Haji, Integrated pumped hydro reverse osmosis system
  optimization featuring surrogate model development in reverse osmosis
  modeling, Applied Energy 352 (2023) 121812.
\newblock \href {https://doi.org/10.1016/j.apenergy.2023.121812}
  {\path{doi:10.1016/j.apenergy.2023.121812}}.

\bibitem{Bilton2011}
A.~M. Bilton, R.~Wiesman, A.~Arif, S.~M. Zubair, S.~Dubowsky, On the
  feasibility of community-scale photovoltaic-powered reverse osmosis
  desalination systems for remote locations, Renewable Energy 36~(12) (2011)
  3246--3256.
\newblock \href {https://doi.org/10.1016/j.renene.2011.03.040}
  {\path{doi:10.1016/j.renene.2011.03.040}}.

\bibitem{DEEP5manual}
{International Atomic Energy Agency}, Deep 5 user manual, 2013.

\bibitem{voutch}
N.~Voutchkov, Desalination project cost estimating and management, CRC Press,
  2019.

\bibitem{316ss}
Shane,
  \href{https://shop.machinemfg.com/comprehensive-guide-to-stainless-steel-pricing-and-grades/}{Comprehensive
  guide to stainless steel pricing and grades}, accessed: 2025-05-14 (March
  2025).
\newline\urlprefix\url{https://shop.machinemfg.com/comprehensive-guide-to-stainless-steel-pricing-and-grades/}

\end{thebibliography}

\appendix
\newpage
\section{Tables of Parameters} \label{app:params}

This appendix contains tables of the parameters used in the model.

\begin{table}[h]
    \centering
    \caption{General Parameters}
    \begin{tabular}{ccc}
        \hline
        \textbf{Parameter} & \textbf{Value} & \textbf{Units} \\
        \hline
        gravitational acceleration & 9.81 & m/s$^2$ \\
        ocean density & 1025 & kg/m$^3$ \\
        distance to shore & 500 & m \\
        ocean temperature & 298.15 & K \\
        water depth & 12 & m \\
        wave direction & 0 \cite{Yu2018} & degrees \\
        
        wave spectrum & Pierson–Moskowitz \cite{Yu2018} & -  \\
        significant wave height & 2.64 \cite{Yu2018} & m \\
        peak period & 9.86 \cite{Yu2018} & s \\

        Fixed Charge Rate & 10.8 \cite{LCOE_DOE} & \% \\
        \hline
    \end{tabular}
    \label{tab:paramsgeneral}
\end{table}

\begin{table}[h]
    \centering
    \caption{SWRO Parameters}
    \begin{tabular}{ccc}
        \hline
        \textbf{Parameter} & \textbf{Value} & \textbf{Units} \\
        \hline
        feedflow total dissolved solids & 35946 \cite{Yu2018} & mg/L \\
        permeate total dissolved solids & 150 \cite{Yu2018} & mg/L \\
        salt molar weight & 58.44 & g/mol \\
        water permeability coefficient & $2.57\times10^{-12}$ \cite{Yu2018} & m$^3$/N-s \\
        recovery ratio & 44.2 \cite{wave} & \% \\
        single SW30HR-380 area & 35 \cite{SW30HR380} & m$^2$ \\
        single SW30HR-380 flow rate & 24.6 \cite{SW30HR380} & m$^3$/day \\
        \hline
    \end{tabular}
    \label{tab:paramsswro}
\end{table}

\begin{table}[h]
    \centering
    \caption{WEC Parameters}
    \begin{tabular}{ccc}
        \hline
        \textbf{Parameter} & \textbf{Value} & \textbf{Units} \\
        \hline
        draft & 9 & m \\
        cg draft factor & -0.7778 & - \\
        unit inertia & 14.57 & m$^2$ \\
        RM5 surface area & 1214 \cite{rm5} & m$^2$ \\
        RM5 flap cost & 3364648.63 \cite{rm5} & \$ \\
        RM5 base cost & 1706415.27 \cite{rm5} & \$ \\
        RM5 bearings cost & 17420.34 \cite{rm5} & \$ \\
        RM5 mooring cost & 997819.2 \cite{rm5} & \$ \\
        RM5 monitoring cost & 616480.27 \cite{rm5} & \$/yr \\
        RM5 marine operations cost & 101387.23 \cite{rm5} & \$/yr \\
        RM5 shore operations cost & 347280.29 \cite{rm5} & \$/yr \\
        RM5 parts cost & 86237.2 \cite{rm5} & \$/yr \\
        RM5 consumables cost & 17480.19 \cite{rm5} & \$/yr \\
        RM5 insurance rate & 2 \cite{rm5} & \% \\
        \hline
    \end{tabular}
    \label{tab:paramswec}
\end{table}

\begin{table}[h]
    \centering
    \caption{PTO Parameters}
    \begin{tabular}{ccc}
        \hline
        \textbf{Parameter} & \textbf{Value} & \textbf{Units} \\
        \hline
        $\ell_2$ & 4.7 & m \\
        $\ell_3$ & 0 & m \\
        max piston stroke & 20 & m \\
        SS316 cost & 2.00 \cite{316ss} & \$/in$^3$ \\
        SS316 density & 0.29 \cite{316ss} & lb/in$^3$ \\
        SS316 yield strength & 206 \cite{316ss} & MPa \\
        SS316 Young's modulus & 164 \cite{316ss} & GPa \\
        cylinder factor of safety & 6 & - \\
        rod factor of safety & 1.5 & - \\
        labor factor & 0.7 & - \\
        end cap attachment factor & 0.3 \cite{ASME_BPVC} & - \\
        cylinder joint efficiency & 0.8 \cite{ASME_BPVC} & - \\
        \hline
    \end{tabular}
    \label{tab:paramspto}
\end{table}

\begin{table}[h]
    \centering
    \caption{Solver Parameters}
    \begin{tabular}{ccc}
        \hline
        \textbf{Parameter} & \textbf{Value} & \textbf{Units} \\
        \hline
        BEM frequencies & 0.2, 0.34, 0.48, ..., 3& rad/s \\
        SysDyn time step & 0.1 & s \\
        SysDyn sim time & 300 & s \\
        \hline
    \end{tabular}
    \label{tab:paramssolve}
\end{table}

\section{Sea States} \label{app:seastates}

This appendix contains a table and a figure of the sea states used in the sensitivity analysis determined through the K-means clustering process described in Section \ref{sec:sensitivity}. 

\begin{table}[h]
    \centering
\caption{Sea states used in sensitivity analysis. The locations column shows how many of the 5 locations the sea state cluster center represents.}
\small
    \label{tab:sea_state_clusters}
    \begin{tabular}{ccc}
    \hline
    \textbf{Peak Period [s]} & \textbf{Significant Wave Height [m]} & \textbf{Locations} \\
    \hline
    13.23112333956173  & 1.7707093836756636 & 2 \\
    10.234793700202095 & 1.4757467428690545 & 2 \\
    22.584459732902232 & 1.1146362808579522 & 2 \\
    13.81664042372318  & 0.8000118789813867 & 2 \\
    10.638863524791551 & 4.635898379970542  & 1 \\
    16.56912770411745  & 2.574394045126751  & 1 \\
    9.191129829212345  & 2.5349366630610004 & 4 \\
    9.976561213254367  & 1.001865613432608  & 5 \\
    5.855232218554694  & 1.0779023831039394 & 4 \\
    13.593014631352435 & 1.1922186822432743 & 4 \\
    12.833294612926467 & 2.063874210861717  & 3 \\
    9.84231198249586   & 1.681761551332142  & 1 \\
    10.284139549631792 & 3.315881129382017  & 2 \\
    13.011144410022865 & 2.732203169230346  & 3 \\
    12.274136752136759 & 6.713700854700875  & 1 \\
    7.900205063645171  & 1.91849027165957   & 5 \\
    17.167737568529763 & 0.9823690678767296 & 1 \\
    7.344880379199283  & 1.3107230273390404 & 5 \\
    8.939529998687709  & 3.193034002888281  & 1 \\
    15.374832971800414 & 3.8642559652928448 & 1 \\
    \hline
    \end{tabular}
\end{table}

\begin{figure}
    \centering
    \includegraphics[width=\linewidth]{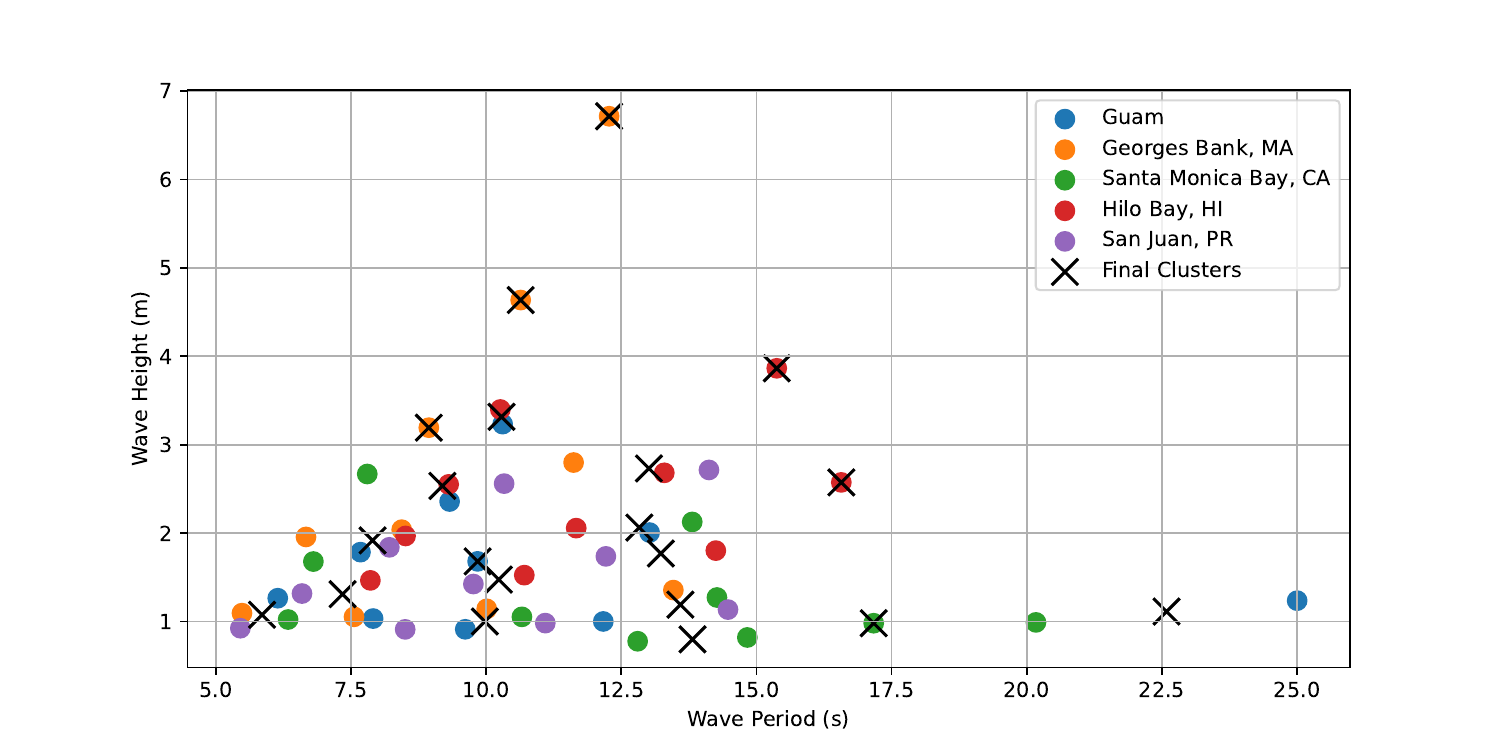}
    \captionof{figure}{Sea states used in sensitivity analysis. Each "X" represents a final cluster center, the points represent the clusters from each buoy, and the color of the points represents the buoy location.}
    \label{fig:sea_states}
\end{figure}

\section{BEM Mesh Discretization} \label{app:mesh}

This section contains a table (Table~\ref{tab:mesh}) showing the mesh discretization used for the BEM hydrodynamics simulation. This discretization was set to ensure the largest panel dimension is less than an eighth of the wavelength at the highest frequency (Table~\ref{tab:paramssolve}). 

\begin{table}[h]
    \centering
    \caption{Mesh discretization for BEM hydrodynamics simulation.}
    \begin{tabular}{cc}
        \hline
        \textbf{Dimension} & \textbf{Number of Panels} \\
        \hline
        Surge/thickness ($t$) & $n_{\text{surge}} = \lceil t \cdot 4/5 \rceil$ \\
        Sway/width ($w$) & $n_{\text{sway}} = \lceil w \cdot 26/30 \rceil$ \\
        Heave/height ($h$) & $n_{\text{heave}} = \lceil h \cdot 8/9.1 \rceil$ \\
        \hline
    \end{tabular}
    \label{tab:mesh}
\end{table}

\section{SWRO Desalination Plant Cost Curves} \label{app:swro_curves}

This appendix contains the coefficients for the CAPEX and OPEX cost curves used in the SWRO plant cost model, which are fitted to curves from Voutchkov’s \emph{Desalination Project Cost Estimating and Management} \cite{voutch}. Each curve is fitted with a power function of the form 
\begin{equation}
    Y = A X^B,
\end{equation}
where $Y$ is either the CAPEX or OPEX, $X$ is the independent variable (either feed flow capacity or permeate flow capacity for CAPEX or average feed flow, average permeate flow, or permeate flow capacity for OPEX), and $A$ and $B$ are the coefficients shown in the tables below. Note that for the HDPE intake pipe costs, the resulting $Y$ from the curve fit is multiplied by the distance to shore (500 m) to get the total cost of the intake pipe. Also note that for the SWRO system costs, since our TDS (35,946 mg/L) lies between the two curves for 35,000 mg/L and 46,000 mg/L, we interpolate between these two curves. Additionally, for the membrane pretreatment costs and the other direct non-energy costs, we use the average of the upper and lower bound curves for our cost model. Lastly, it is important to note that all costs in the following tables are in 2018 USD, since that is the year of the publication of Voutchkov’s \emph{Desalination Project Cost Estimating and Management} \cite{voutch}, and the cost model is fitted to curves from that book. The SWRO cost model used in this study multiplies the resulting costs from these curves by an inflation factor (1.26) to convert them to 2025 USD.

\begin{table}[h!]
\centering
\tiny
\begin{tabular}{llcccc}
\hline
\textbf{Figure} & \textbf{Curve} & \textbf{A} & \textbf{B} & \textbf{X units} & \textbf{Y units} \\
\hline
4.3  & HDPE Offshore Intake & 0.001792 & 0.7837 & feed m$^3$/day & \$K/m \\
4.5  & Intake screens - band screens & 0.007936 & 1.0210 & feed m$^3$/day & \$K \\
4.6  & Bulk filtration - wedgewire screens & 0.04816 & 0.8412 & feed m$^3$/day & \$K \\
4.7  & Intake screens - microscreens & 0.06158 & 0.8466 & feed m$^3$/day & \$K \\
4.10 & Membrane pretreatment costs - Upper Bound & 1.0289 & 0.8127 & feed m$^3$/day & \$K \\
4.10 & Membrane pretreatment costs - Lower Bound& 0.7656 & 0.7904 & feed m$^3$/day & \$K \\
4.13 & Single pass SWRO system - Feed TDS = 46,000 mg/L & 4.9006 & 0.7925 & permeate m$^3$/day & \$K \\
4.13 & Single pass SWRO system - Feed TDS = 35,000 mg/L & 5.0617 & 0.7779 & permeate m$^3$/day & \$K \\
4.18 & Stabilization post treatment - Lime-CO$_2$ system & 6.0711 & 0.6024 & permeate m$^3$/day & \$K \\
4.19 & Post treatment disinfection - bulk sodium hypochlorite & 0.4992 & 0.6000 & permeate m$^3$/day & \$K \\
4.18 & Stabilization post treatment - Calcite-CO$_2$ system & 3.2145 & 0.6026 & permeate m$^3$/day & \$K \\
\hline
\end{tabular}
\caption{CAPEX cost curves, fitted to curves from Voutchkov’s \emph{Desalination Project Cost Estimating and Management}, Chapter 4 \cite{voutch}. The Figure column refers to the figure in Voutchkov’s book that the curve is fitted to. Note that all costs in this table are in 2018 USD.}
\end{table}

\begin{table}[h!]
\centering
\tiny
\begin{tabular}{llcccc}
\hline
\textbf{Figure} & \textbf{Curve} & \textbf{A} & \textbf{B} & \textbf{X units} & \textbf{Y units} \\
\hline
5.3  & Offshore intake - HDPE pipe & 0.0136 & 0.7804 & feed m$^3$/day & \$ /m/year \\
5.5  & Intake screens - band screens & 0.0002724 & 1.0227 & feed m$^3$/day & \$K/year \\
5.6  & Bulk filtration - wedgewire screens & 0.001959 & 0.8430 & feed m$^3$/day & \$K/year \\
5.7  & Intake screens - microscreens & 0.002714 & 0.8451 & feed m$^3$/day & \$K/year \\
5.11 & Membrane pretreatment costs - upper & 0.04874 & 0.8139 & feed m$^3$/day & \$K/year \\
5.11 & Membrane pretreatment costs - lower & 0.05010 & 0.7877 & feed m$^3$/day & \$K/year \\
5.12 & SWRO system - TDS 46,000 mg/L & 0.2098 & 0.7922 & permeate m$^3$/day & \$K/year \\
5.12 & SWRO system - TDS 35,000 mg/L & 0.1969 & 0.7814 & permeate m$^3$/day & \$K/year \\
5.15 & Stabilization - Lime-CO$_2$ system & 0.6040 & 0.5993 & production capacity m$^3$/day & \$K/year \\
5.16 & Disinfection - sodium hypochlorite & 0.01355 & 0.7804 & production capacity m$^3$/day & \$K/year \\
5.17 & Other direct non-energy costs - upper & 0.3652 & 0.7517 & permeate m$^3$/day & \$K/year \\
5.17 & Other direct non-energy costs - lower & 0.0329 & 0.7819 & permeate m$^3$/day & \$K/year \\
5.18 & Indirect O\&M Costs - upper & 0.3777 & 0.7491 & permeate m$^3$/day & \$K/year \\
5.18 & Indirect O\&M Costs - lower & 0.1685 & 0.7373 & permeate m$^3$/day & \$K/year \\
5.15 & Stabilization - Lime-CO$_2$ (alt) & 0.3411 & 0.5996 & production capacity m$^3$/day & \$K/year \\
\hline
\end{tabular}
\caption{OPEX cost curves, fitted to curves from Voutchkov’s \emph{Desalination Project Cost Estimating and Management}, Chapter 5 \cite{voutch}. The Figure column refers to the figure in Voutchkov’s book that the curve is fitted to. Note that all costs in this table are in 2018 USD.}
\end{table}
\end{document}